\documentclass[8.5pt,twoside,twocolumn]{article}
\usepackage[super,sort&compress,comma]{natbib} 
\usepackage{mhchem}
\usepackage{times,mathptmx}

\usepackage{sectsty}
\usepackage{balance} 
\usepackage{graphicx} 
\usepackage{lastpage}
\usepackage[format=plain,justification=raggedright,singlelinecheck=false,font=small,labelfont=bf,labelsep=space]{caption} 
\usepackage{fancyhdr}

\usepackage[english]{babel}
\usepackage{xspace}
\usepackage{color}
\newcommand{\dx}{d\mathbf{x}}
\newcommand{\vv}[1]{\mathbf{#1}}
\newcommand{\vx}{\left(\vv{x}\right)}
\newcommand{\vxt}{\left(\vv{x},t\right)}
\newcommand{\grad}{{\mathbf{\nabla}}}

\newcommand{\prhopt}{\ensuremath{ {\partial \rho_p \over \partial t} } }
\newcommand{\rhox}{\ensuremath{ \rho_p\left( \vv{x} \right) } }
\newcommand{\rhoxt}{\ensuremath{ \rho_p\left( \vv{x},t \right) } }
\newcommand{\rhoxe}{\ensuremath{ \rho_p^{e}\left( \vv{x} \right) } }
\newcommand{\rhor}{\ensuremath{ \rho_p\left( r \right) } }
\newcommand{\rhort}{\ensuremath{ \rho_p\left( r,t \right) } }
\newcommand{\prhort}{\ensuremath{ {\partial \rho_p\left( r,t \right)  \over \partial r} }}
\newcommand{\rhore}{\ensuremath{ \rho_p^{e}\left( r \right) } }
\newcommand{\muxt}{\ensuremath{ \mu_p\left( \vv{x},t \right) } }

\newcommand{\rhop}{\ensuremath{ \rho_p}}
\newcommand{\aaa}{\ensuremath{ \alpha}}
\newcommand{\FF}{\mathcal{F}}

\newcommand{\Rg}{R_{gel}}

\newcommand{\kb}{k_\mathrm{B} T}
\newcommand{\bbb}{\beta}

\begin{document}

\thispagestyle{plain}
\fancypagestyle{plain}{
\fancyhead[L]{\includegraphics[height=8pt]{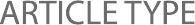}}
\fancyhead[C]{\hspace{-1cm}\includegraphics[height=20pt]{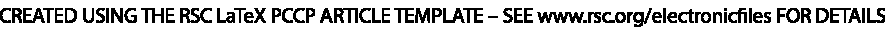}}
\fancyhead[R]{\includegraphics[height=10pt]{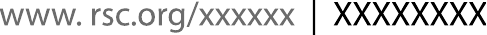}\vspace{-0.2cm}}
\renewcommand{\headrulewidth}{1pt}}
\renewcommand{\thefootnote}{\fnsymbol{footnote}}
\renewcommand\footnoterule{\vspace*{1pt}%
\hrule width 3.4in height 0.4pt \vspace*{5pt}} 
\setcounter{secnumdepth}{5}

\makeatletter 
\def\subsubsection{\@startsection{subsubsection}{3}{10pt}{-1.25ex plus -1ex minus -.1ex}{0ex plus 0ex}{\normalsize\bf}} 
\def\paragraph{\@startsection{paragraph}{4}{10pt}{-1.25ex plus -1ex minus -.1ex}{0ex plus 0ex}{\normalsize\textit}} 
\renewcommand\@biblabel[1]{#1}            
\renewcommand\@makefntext[1]%
{\noindent\makebox[0pt][r]{\@thefnmark\,}#1}
\makeatother 
\renewcommand{\figurename}{\small{Fig.}~}
\sectionfont{\large}
\subsectionfont{\normalsize} 

\fancyfoot{}
\fancyfoot[LO,RE]{\vspace{-7pt}\includegraphics[height=9pt]{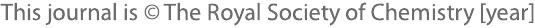}}
\fancyfoot[CO]{\vspace{-7.2pt}\hspace{12.2cm}\includegraphics{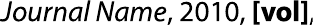}}
\fancyfoot[CE]{\vspace{-7.5pt}\hspace{-13.5cm}\includegraphics{RF}}
\fancyfoot[RO]{\footnotesize{\sffamily{1--\pageref{LastPage} ~\textbar  \hspace{2pt}\thepage}}}
\fancyfoot[LE]{\footnotesize{\sffamily{\thepage~\textbar\hspace{3.45cm} 1--\pageref{LastPage}}}}
\fancyhead{}
\renewcommand{\headrulewidth}{1pt} 
\renewcommand{\footrulewidth}{1pt}
\setlength{\arrayrulewidth}{1pt}
\setlength{\columnsep}{6.5mm}
\setlength\bibsep{1pt}

\twocolumn[
  \begin{@twocolumnfalse}
\noindent\LARGE{\textbf{ Dynamic density functional theory of protein adsorption on polymer-coated nanoparticles }}
\vspace{0.6cm}

\noindent\large{\textbf{Stefano Angioletti-Uberti,$^{\ast}$\textit{$^{a,b,c}$} Matthias Ballauff,\textit{$^{b,c}$} and Joachim Dzubiella \textit{$^{b,c}$}}}\vspace{0.5cm}

\noindent\textit{\small{\textbf{Received Xth XXXXXXXXXX 20XX, Accepted Xth XXXXXXXXX 20XX\newline First published on the web Xth XXXXXXXXXX 200X}}}

\noindent \textbf{\small{DOI: 10.1039/b000000x}}
\vspace{0.6cm}

\noindent \normalsize{We present a theoretical model for the description of the adsorption kinetics of globular proteins onto charged 
core-shell microgel particles based on Dynamic Density Functional Theory (DDFT).
This model builds on a previous description of protein 
adsorption thermodynamics [Yigit \textit{et al}, Langmuir 28 (2012)], shown to well interpret 
the available calorimetric experimental data of binding isotherms. In practice, a spatially-dependent free-energy functional 
including the same physical interactions is built, and used to study the kinetics via a generalised
diffusion equation. 
To test this model, we apply it to the case study of Lysozyme adsorption on PNIPAM 
coated nanoparticles, and show that
the dynamics obtained within DDFT is consistent with that extrapolated from experiments. 
We also perform a systematic study of the effect of various parameters in our model, and investigate 
the loading dynamics as a function of proteins' valence and hydrophobic adsorption energy, 
as well as their concentration and that of the nanoparticles.
Although we concentrated here on the case of adsorption for a single protein type, 
the model's generality allows to study multi-component system, providing a reliable 
instrument for future studies of competitive and cooperative adsorption effects often 
encountered in protein adsorption experiments.}
\vspace{0.5cm}
 \end{@twocolumnfalse}
  ]


\footnotetext{\textit{$^{aa}$~E-mail: sangiole@physik.hu-berlin.de}}
\footnotetext{\textit{$^{a}$~Institut f{\"u}r Physik, Humboldt-Universit{\"a}t zu Berlin,  12489 Berlin, Germany;}}
\footnotetext{\textit{$^{b}$~Soft Matter and Functional Materials, Helmholtz Zentrum Berlin, 14109 Berlin, Germany}}



\section{Introduction \label{sec:intro}}
Protein adsorption on various materials is a fascinating problem with important repercussions for the
development of a large number of diverse technologies. These include food manufacturing processes, 
biomaterials for medical implants and functionalised nanoparticles for targeted drug delivery, 
among many others \cite{biotech-review}.
The need to understand protein adsorption arises from the fact that the characteristics of the 
protein layer formed upon adsorption (often called the ``protein corona'' in the case of nanoparticles), 
dictates the subsequent interaction of the material with biological entities, 
for example bacteria, antibodies or cells \cite{nature-protein-corona, nature-protein-corona2}. 
Hence, depending on the type of application, one would typically either prevent protein 
sorption altogether or to allow for some selectivity in the process. 
In this regard, polymer coatings have been shown to represent a viable way to control 
protein adsorption, and their intense study gave rise to a vast literature which would be 
impractical to recapitulate here. The interested reader is referred to a very recent review
of the subject by Haag \textit{et al} \cite{protein-critique}, whereas here we will only 
briefly discuss previous theoretical approaches aimed at describing protein adsorption kinetics.\\*
\begin{figure*}[h!]
\center
\includegraphics[width=0.48\textwidth]{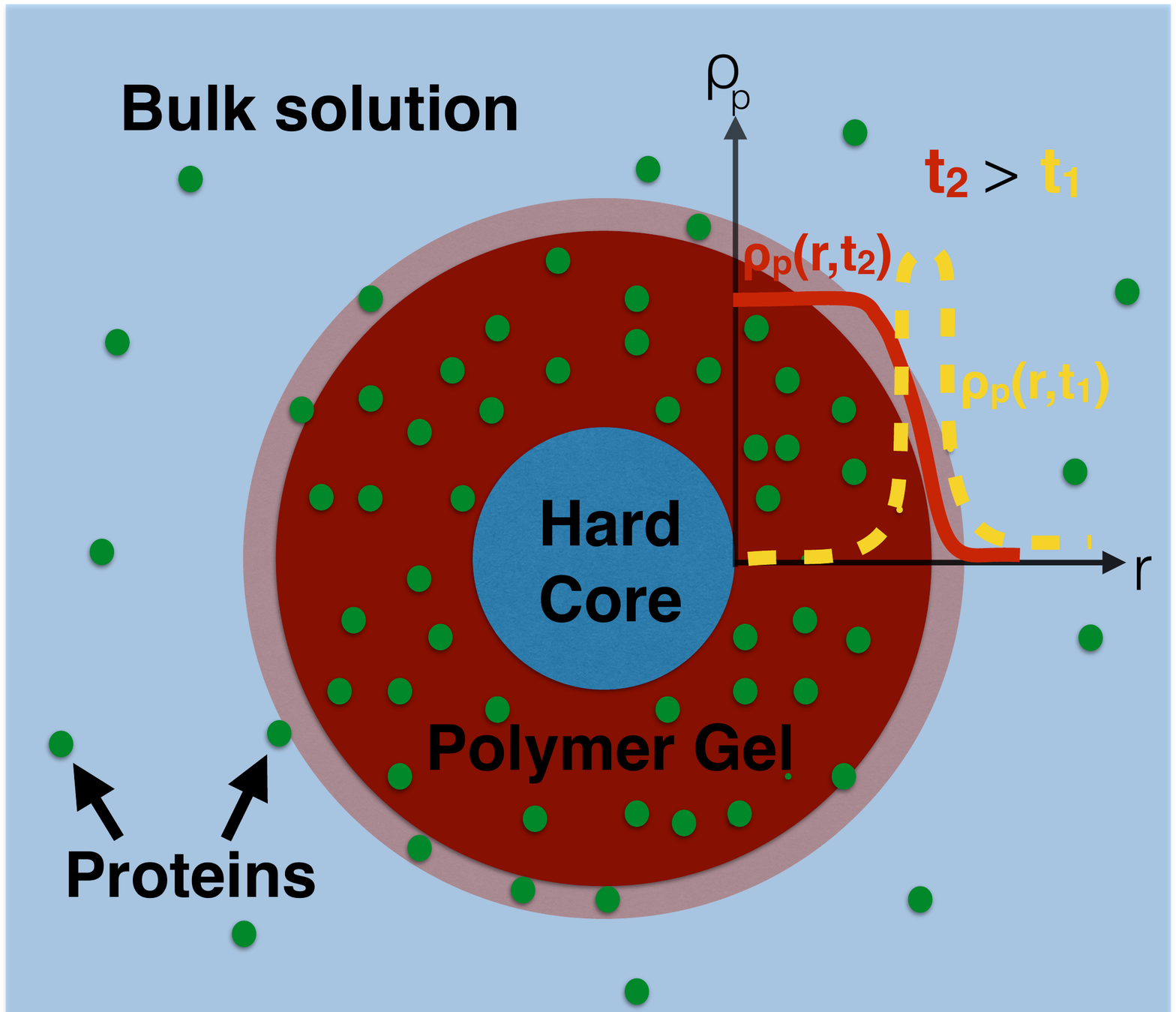}
\caption{Schematic representation of our system. A $60$~nm hard-core PMMA nanoparticle (blue) coated with a 
cross-linked polymer network (hydrogel) of PNIPAM $90$~nm thick (dark red) with an interface width of around $10~$nm (light red). 
All nanoparticle dimensions are scaled with the correct size ratio (protein are represented larger than their actual size). 
This core-shell nanoparticle, which we also refer to as nanogel, is immersed in a protein solution (green points). 
Proteins are described in our DDFT model as a continuous, time-dependent radial density field $\rho_p(r,t)$ with origin at the nanoparticle's hard 
core / polymer boundary. Here, two density profiles corresponding to different times (yellow dashed line and, at later times, red, continuous line)
are shown. Within this model, mixture of different protein types can also be easily treated.
}
\label{fig:system-schematic}
\end{figure*}
From a theoretical point of view, protein adsorption kinetics has been mainly studied based 
on three different approaches: ideal diffusion equations \cite{lysozime-soft-matter}, 
Langmuir-type models (also called mass-balance equations) \cite{linse, riviere}, and models
based on a "generalised diffusion approach", also termed "molecular approach" 
\cite{fang-szleifer,fang-szleifer-charge,szleifer-carignano,fang-szleifer-compare}.
%
%
Given their very nature, models based on \textit{ideal} diffusion cannot 
capture the complex dynamics of protein adsorption since all the important interactions 
between proteins and their environment are completely neglected. 
For this reason, these models do not reproduce at long timescales the right 
thermodynamics, which is a crucial ingredient to obtain the kinetics, 
as well as for physical consistency. In fact, as we will show later,
calculations based on ideal diffusion produce loading timescales estimates which 
can be off by two orders of magnitude from those deduced from experiments,
although fortuitous cancellation of errors can sometime occur partially correcting
the problem in certain cases (see Sec.~\ref{sec:debye}). For this reason, care should be taken to avoid 
over-interpretation of experimental observations based on these simple theoretical description, 
in particular regarding the proteins' mobility \cite{lysozime-soft-matter}.
Despite this caveat, not only protein adsorption but also drug loading and release dynamics onto and from 
nanoparticles have been typically discussed based on these 
simple models \cite{drug-release1,drug-release2}.\\
%
%
Langmuir models by construction give the correct thermodynamics of protein adsorption. 
This is often sufficient to correctly reproduce the observed dynamics when 
single-type protein adsorption occurs and adsorption relies on the Langmuir picture of independent, 
single binding sites without collective or cooperative effects. 
However, when multiple protein types coexist, it is hard to guess a \textit{a priori} the validity of these
assumptions or whether more complex interactions occur. For example, mutual interactions 
between proteins can induce cooperative adsorption that cannot be casted in terms of single, independent binding sites.
Quite generally, it is not possible to say if intermediate, metastable adsorption states observed 
in protein adsorption, are correctly described by these models.\\
Finally, one important information one would like to have access to is the 
full density profile as a function of time, not just the amount of adsorbed protein as in a Langmuir model. 
These profiles can be highly inhomogeneous, in particular for multiple-component systems, and vary strongly
in time. Since it is the outer protein shell in contact with the biological environment that determines a nanoparticle's
interaction, a correct description of such inhomogeneities  is important to understand its functional behaviour. 
For these reasons, we choose to use a general microscopic approach, as pioneered by Szleifer 
and coworkers, who built several models to study protein adsorption for various types of both coated 
and bare infinite planar surfaces \cite{fang-szleifer,fang-szleifer-charge,szleifer-carignano,
fang-szleifer-compare}. 
Our model is similar to the latter in the sense that we start from the same theoretical 
framework, i.e. Dynamic Density Function Theory (DDFT). 
However, apart from studying protein adsorption on curved, finite systems like nanoparticles rather 
than planar surfaces, we will combine DDFT with a different free-energy functional. The latter
was inspired by the work of Yigit \textit{et al.} \cite{joe-langmuir} who proposed a coarse-grained model
that was shown to well described the protein adsorption for our system. In particular, it included electrostatic cooperativity 
effects due to the changing net charge of the hydrogel by increasing protein adsorption. Furthermore, it 
 demonstrated that Langmuir models are equivalent to more general description in terms
of excluded volume packing effects in the limit of low protein packing fractions in the gel. 
The latter finding relieves us from the assumption of independent, single binding sites and allows us to 
describe protein adsorption (especially of multicomponent mixtures) in a more versatile way based on packing effects. 
As in Ref.~\cite{joe-langmuir}, we include here the electrostatic contributions
within an effective description based on the concept of the Donnan potential. 
The advantage of this  treatment allow us to clearly separate \textit{global} electrostatic effects 
from \textit{specific}, i.e. protein-dependent ones, shedding some light on the magnitude
and relative importance of each of them in different scenarios.\\*
The remaining of the paper is structured as follows. In Sec.~\ref{sec:theory} we first give a brief, heuristic
introduction to the basic DDFT equations, and then proceed to explain the details of our model trying
to clearly state all its underlying assumptions and their validity. 
In Sec.~\ref{sec:compare}, before we procede to describe the DDFT results, 
we discuss two analytically solvable models based on the ideal diffusion equation
to obtain a first, rough estimate of the timescales expected to appear in our system. 
Sec.~\ref{sec:numerical-results} reports our numerical results for the case of Lysozyme 
adsorption on PNIPAM coated nanogels, and compare them to extrapolation from
the available experimental data as well as those obtained from the solution of the ideal 
diffusion equation for the same system. 
We also report a systematic analysis of the role of various interactions
and parameters of our model, and critically discuss the obtained results.
Finally, we draw our conclusion in Sec.~\ref{sec:conclusion}.

\section{Theoretical Model \label{sec:theory} }

\subsection{A short introduction to DDFT  \label{subsec:ddft} }

At its root, DDFT is nothing but a generalised diffusion 
equation describing the density evolution of out-of-equilibrium systems undergoing Brownian dynamics 
\cite{marconi-tarrazona, lowen-ddft1,lowen-ddft2, lowen-ddft3}. 
Although a formal derivation starting from the Smoluchowski 
equation can be built \cite{ddft-formal, marconi-tarrazona}, 
a less rigorous but more intuitive heuristic argument can be given \cite{dft-ddft-review}, which we will outline here for simplicity. 
We start with the continuity equation:

\begin{equation}
\prhopt = -\nabla\cdot \vv{J}_p
\label{eq:continuity}
\end{equation}

where \rhoxt is the space and time-dependent density field of specie $p$ and $\vv{J}_p$ (also
a function of time and space) its associated flux.
We \textit{assume} $\vv{J}_p$ to be linear in the gradient of the chemical potential of the same
specie, $\mu_p$, scaled by the inverse temperature $\beta = 1/k_{\mathrm{B}} T$ (where $T$ is the 
absolute temperature and $k_{\mathrm{B}}$ is Boltzmann's constant), i.e. formally:

\begin{align}
\vv{J}_p\vxt = - D_p(\vv{x})\rhoxt \grad \beta\mu\vxt.
\label{eq:flux}
\end{align}

The linearity coefficient in Eq.~\ref{eq:flux} is nothing but the diffusion coefficient $D_p$.
Plugging Eq.~\ref{eq:flux} into Eq.~\ref{eq:continuity} we obtain a "generalised diffusion equation"

\begin{align}
\prhopt = \nabla \cdot D_p \rhop \nabla\beta\mu_p
\label{eq:diffusion}
\end{align}

which can be written in a more insightful form by splitting the chemical
potential into ideal and excess contribution, $\mu^{id}$ and $\mu^{exc}$, giving:

\begin{align}
\prhopt &= \nabla\cdot  D_p \left[ \beta \mu_p^{id} + \beta\mu_p^{exc} \right] \nonumber \\
	  &=  \nabla\cdot D_p \grad \rho_p + \nabla \cdot D \rho_p \grad \bbb\mu_p^{exc}, 
\label{eq:diffusion2}
\end{align}

where in the last line we have made the substitution $\beta\mu_p^{id} = \log \rho_p / \rho_0$, $\rho_0$ being 
a reference density which we fix to the standard molar density of $1$~M.
The first term on the r.h.s of Eq.~\ref{eq:diffusion2} is the ideal diffusion term, which tends to 
smoothen any possible density gradient within the system.
If no inter-particle interactions nor any external field were present,the excess term would be zero.
With the additional constraint of a constant diffusion coefficient $D_p$, one would then recover 
the well known formula $\prhopt = D_p \nabla^2 \rho_p $, i.e. the ideal diffusion equation. 
In the general, more realistic case, $\mu^{ex} \neq 0$ and we need a way to calculate this term
to determine the dynamical behaviour of the system.\\
This is provided by classical, equilibrium DFT \cite{evans,hansen-macdonald}, which 
gives the following expression for the chemical potential: 

\begin{align}
\mu_p = {\delta \FF[ \{\rho_p \} ] \over \delta \rho_p},
\label{eq:chemPot}
\end{align}

where $\FF[ \{\rho_p \} ]$ is the free-energy functional of our system, which depends
on the densities of all species (labelled by the subscript $p$).\\
The underlying assumption at the basis of DDFT is that Eq.~\ref{eq:chemPot}, 
remains valid also out of equilibrium, i.e. one is under quasi-equilibrium conditions. 
A quasi-equilibrium assumption is already implicit in writing Eq.~\ref{eq:flux} as 
the gradient of a chemical potential, implying the presence of a conservative field, 
whereas under full non--equilibrium conditions the true force might be non-conservative.
For our specific system, this requires that all other degrees 
of freedom like the density field of ions and solvent molecules quickly relax around 
the instantaneous "equilibrium" configuration of the protein density. Moreover, the 
frequency of external time-dependent fields should not be comparable to 
the typical relaxation frequency of the system. 
In these latter scenarios, more complex theories have to be used, such as the 
recently developed Power Functional Theory of Schmidt and Brader 
\cite{power-functional1,power-functional2}.\\
When the underlying approximations are met, the agreement between theory and experiments or numerical 
Brownian dynamics simulations is excellent. In this regard, DDFT has proven to be a versatile 
instrument, allowing to describe a large variety of phenomena, ranging from the sedimentation 
of colloids under gravity \cite{ddft-sedimentation, archer2,krueger} and colloidal dynamics in polymers 
mixture \cite{ddft-colloid-polymer} to the dewetting of evaporating nanoparticle films \cite{archer1},
or the kinetics of colloids diffusing in confined geometries \cite{ddft-narrow-channels, rauscher1}. 
As we are about to show in the later sections, protein adsorption kinetics on polymer-coated charged nanoparticles 
also appears to be treatable within this framework.\\
\subsection{A free-energy functional for protein adsorption on charged nanogels \label{sec:building-functional}}

As implied by Eq.~\ref{eq:chemPot}, in order to treat our problem using DDFT we need to
specify the free-energy functional for our system $\FF[\{\rho_p\}]$. 
In its most general form, for \textit{any} classical system $\FF$ can be written as :

\begin{align}
\FF =& \FF[\{\rho_p\}]  = \FF^{id} + \FF^{ext} + \FF^{exc} \nonumber  \\
   =& \sum_{p} \int_V\kb\rhox \left[ \ln\left({\rhox \over \rho_0 }-1\right)\right] \dx + \nonumber \\
     & \sum_{p} \int_V V^{ext} \rhox  \dx +  \FF^{exc}[\{\rhox\}]. 
\label{eq:Ftotal}
\end{align}

where the sum is over all $p$ species and the integral has to be read as a three-dimensional integral over the
whole volume $V$. Although we will not always make it explicit in the notation, it should be reminded that~$\rho_p$ 
and all other quantities depending on it are both space \textit{and} time-dependent quantities.
The first term in Eq.~\ref{eq:Ftotal} is the free-energy density for an ideal gas of particles, 
the second describes the coupling between the density and an external potential $V^{ext}$ and 
the third, typically called the $excess$ functional, describes inter-particles interactions.\\
No \textit{exact} form exists for $\FF^{exc}$, hence Eq.~\ref{eq:Ftotal}
just shifts the problem from the definition of $\FF$ to that of $\FF^{exc}$. 
However, one should notice that in many cases not only most of the free-energy 
contribution is accounted for by the first two terms, but also that a few useful 
approximations exist for $\FF^{exc}$, depending on the type of system under consideration.
Among these approximations, the simplest possible one, which will also be employed here, is the so-called Local 
Density Approximation (LDA). In the LDA,  one assumes that the excess free-energy density per particle at a point $\vv{x}$ is a 
function of the local density at $\vv{x}$ only, and equal to its value for an homogeneous system at the same density, 
$\epsilon^{exc}(\{ \rho_p \}$, i.e. 

\begin{equation}
\FF^{exc} = \sum\limits_p \int_V \epsilon^{exc}(\{ \rhox \})\rhox \dx.
\label{eq:lda-approx}
\end{equation}

If density fluctuations occur on a scale that is large compared to the interaction range of the particles, 
each of them ``feels'' around it an homogeneous environment, and the system should be well described by 
the LDA. When this is not the case, one can resort to more complex non-local functionals, e.g. those based on a 
mean-field \cite{hansen-macdonald,joe-likos} or ``weighted density'' approximation \cite{weighted-density}.\\*
%
The crucial step in defining our model for protein adsorption is the correct description of 
the important physical forces that play a role in the adsorption process. 
In practice, this translates into finding a good approximation for the 
free-energy functional $\FF[\left\{\rho\right\}]$. 
In doing so, we will keep in mind that an important quality we would like to endow our functional with 
is to contain only experimentally accessible quantities. This latter property will
allow us to make direct contact with experiments, which eventually represent the most
important test for the validity of our theory.

%

Instead of trying to build a general model, we focus here on describing the case of protein adsorption on charged 
hydrogel-coated nanoparticles (which we sometimes refer to as nanogels). 
For this type of system, which still represent a broad category of important experimental cases,
we show here how a simplified but robust model can be built by including a coarse-grained 
description of the major physical forces playing a role in 
the adsorption process, throwing out less relevant details and keeping all functional forms as 
simple as possible.
For example, for the small but finite concentration of proteins found in these nanoparticles, 
the most relevant information about protein-protein interactions is well captured by a measurable 
thermodynamic quantity such as the second virial coefficient. Clearly, by using this parameter 
as a proxy for the full interaction potential we are making assumptions that restrict the validity of 
the model, which however remains general enough to be applicable to the majority of cases we 
would like to describe. In practice, we pay in generality what we get back in 
reliability and usability of the model.\\*
Based on similar premises, Yigit \textit{et al} presented in Ref.~\cite{joe-langmuir} a minimal
thermodynamic model for protein adsorption onto charged nanoparticles that was shown to well compare with many
available experimental data. For this reason, we decided to build our DDFT model by including 
the same terms. Hence, the free-energy functional we propose is the following:

\begin{align}
\FF &= \FF^{id} + \FF^{ext} + \FF^{exc} \nonumber \\
   &= \FF^{id} + \left( \FF^{ads} + \FF^{electro} \right) + \FF^{exc} \nonumber \\
   &= \FF^{id} + \FF^{ads} + \FF^{\mathrm{Born}} + \FF^{\mathrm{Don}} + \FF^{exc}\nonumber \\
   &= \sum\limits_{p} \int_V\kb\rhox \left[ \ln\left({\rhox \over \rho_0 }-1\right)\right] \dx \nonumber \\
   &+ \int_V\rhox V^{ads}\left(\vv{x}\right) \dx - \int_V \rhox V^{\mathrm{Born}}\left(\vv{x}\right) \dx \nonumber \\
   &+ \int_V z_p \rhox V^{\mathrm{Don}}\left[ \{\rhox\}^* \right ]\dx \nonumber \\
   &+\int_V \rhox \epsilon^{exc}\left(\{\rhox\}\right) \dx,
\label{eq:Ftotal-decomposed}
\end{align}

where the asterisk in the definition of $V^{\mathrm{Don}}$ means that when calculating its contribution to 
the chemical potential $\mu_p$ by taking the functional derivative, this should be done 
at a fixed value of $V^{\mathrm{Don}}$ to properly account for the charge-neutrality condition.\\
The first term in Eq.~\ref{eq:Ftotal-decomposed} $\FF^{id}$ is the ideal gas term. It accounts for the 
translational free-energy (entropy) of proteins in solutions.
As previously explained, taken alone this term gives rise to the ideal diffusion equation.
The remaining terms are instead due to interactions within the system. 
Two of them, $\FF^{ads}$ and $\FF^{electro}$, depend on the protein-nanogel interaction, 
whereas $\FF^{exc}$ accounts for protein-protein interactions.\\*
%
%
$\FF^{ads}$ measures the intrinsic adsorption free-energy arising from protein-specific forces
between proteins and the gel, such as hydrophobic and hydration forces or salt-bridges \cite{jackson}.
We model this term as simply as possible using:

\begin{align}
\FF^{ads} = &\int_V\rhox V^{ads} \dx\\
V^{ads}(r) = & S(r) \Delta G^{ads}\\
S(r) = & \left[1 - \mathrm{Fe}( r, \Rg, \sigma)\right].
\label{eq:switch}
\end{align}

Here, $\Delta G^{ads}$ is the intrinsic adsorption energy per protein and $S$ a switching function, describing 
the change of environment from that of the bulk gel to that of the bulk protein solution, where 
$\textrm{Fe}( r, \mu, \aaa) = 1 / ( 1 + \exp[( r - \mu)/\aaa] ) $ is the Fermi function with 
inflection point at $\mu$ and width $\aaa$, and $r=\mid\vv{x}\mid$ measures the distance from the centre
of the nanoparticle.
This choice of $S$ ensures that the intrinsic interactions are local and present only when the 
protein effectively enters in the gel. A finite value for $\sigma$ also implies that the gel-bulk 
solution boundary is not atomically sharp but varies within a distance $\sigma$ of a few nanometers, 
comparable to the average cross-linking distance typically found in the polymer network of this system. 
For this reason, and to maintain consistency, the same type of spatial dependence 
is chosen also for the gel density and the protein's diffusion coefficient (which is a space dependent quantity varying between the bulk 
solution and the gel matrix), i.e.:

\begin{align}
\rho_{gel}(r) = & \rho_{gel}^{bulk} \left[1-S\left(r\right)\right]  \label{eq:rhogel} \\
D_p(r) = & D_p^{gel} + ( D_p^{bulk} - D_p^{gel} ) S(r) \label{eq:Dcoeff}
\end{align}

where $D_p^{bulk}$  and $D_p^{gel}$ are the protein diffusion coefficient in the bulk solution and in the polymer gel, 
respectively \cite{lysozime-diffusion,soft-matter-kinetics}, and $\rho_{gel}^{bulk}$ is the polymer bulk number density. Other choices for these profiles with similar, 
physically justified  shapes can be considered without affecting the simulation result.\\*
The electrostatic free-energy $\FF^{electro}$ is purely dictated by the charge of the protein and the 
nanogel, which in turn depend on the pH of the system as well as salt concentration and can be further split into two terms, $\FF^{\mathrm{Don}}$ and $\FF^{\mathrm{Born}}$. 
$\FF^{\mathrm{Don}}$ is an electrostatic contribution due to the difference in the electrostatic potential between the gel and bulk solution.
This so-called Donnan potential, derived by imposing local charge neutrality in the system \cite{manning-donnan,joe-langmuir}, 
depends on both the fixed charges of the nanogel as well as the mobile proteins and salt ions. 
The explicit form of the Donnan potential is: 

\begin{align}
e\beta V^{\mathrm{Don}}\vx &= {\tilde{V}^{\mathrm{Don}}}\vx = \ln\left[   \sqrt{ y\vx^2 + 1} + y\vx \right], \hspace{0.2cm}\mathrm{with} \label{eq:donnan} \\
y\vx &= { z_{gel} \rho_{gel}^c\vx + \sum\limits_p z_p\rhox \over z_{s}\rho_{s}^{bulk} } \label{eq:donnan2}
\end{align}

where $\rho_{gel}^c\vx$ and $z_{gel}$ are the number density of charged monomers (i.e. $\rho_{gel}^c=f_c \rho_{gel}\vx$, where $f_c$ is
the fraction of charged monomers) and the monomer charge, respectively. Correspondingly, 
$\rho_{s}^{bulk}$ and $z_{s}$ are the bulk concentration of salt and the charge of 
a salt ion and finally $z_p$ is the charge of a protein of type $p$.%

In principle, one could calculate the full electrostatic energy of the system by building a 
density functional that includes also the densities of salt ions. However, the size of these ions is
much smaller than that of a protein, hence they are a lot faster. 
This allows to assume that they are in local equilibrium with the density of the ``slow'' charges, 
those of the proteins and the gel. This separation of timescales greatly reduces the computational 
complexity of the problem \cite{fang-szleifer}, and the electrostatic contributions can be efficiently calculated.
One way to do this would be to fully solve the underlying Poisson-Boltzmann equations,
at a fixed charge density given by the instantaneous realisation of the protein density field.
However, if one coarse-grains the system on distances larger than the Debye screening length, 
a more efficient approach is to simply assume local charge neutrality, as we do here. 
With this choice, in the bulk of the gel we recover exactly the same value of the electric field obtained 
solving the Poisson-Boltzmann equation. Moreover, we recall that in our model all local 
properties including the electrostatic potential change from that of the gel to their 
bulk solution value within a distance of $\sigma$ from the gel boundary (Eq.~\ref{eq:switch}). 
Since our choice for $\sigma$ is close to the Debye screening 
length $\ell_{Debye}$ ($\approx 3.6~nm$ at the salt concentrations considered here), 
our minimal model is also in semi-quantitative agreement with the Poisson-Boltzmann solution
for the \textit{variation} of the electrostatic field at the gel-bulk solution interface.\\*
%
%
The second term in the electrostatic energy is the Born transfer energy $F^{\mathrm{Born}}$, 
which simply describes the change in the self-energy of the charged proteins 
due to the different screening properties in the gel matrix and the bulk solution, whose known form is 
\cite{jackson}:

\begin{align}
\beta V^{\mathrm{Born}}\left(\vv{x}\right) =  & { z_p^2 l_{B} \over 2 r_p } { \kappa \vx r_p \over \left( 1 + \kappa \vx r_p \right )} \\
\kappa\vx = & \sqrt{ 4 \pi \lambda_B \rho_{local} \vx} \\
		     = & \sqrt{ 4 \pi \lambda_B \left(\rho_{gel}^c\vx+\rho_{s}\vx\right)}  \nonumber \\
\rho_{s}\vx = & \rho_s^{bulk} \left( e^{\left(-z_s \tilde{V}^{\mathrm{Don}\vx}\right)} + e^{\left(+z_s \tilde{V}^{\mathrm{Don}\vx}\right)} \right) 
\label{eq:born}
\end{align}

where $\lambda_B = {e^2 \over 4 \pi \epsilon_0\epsilon  k_{\mathrm{B}} T }$ is the Bjerrum length (taken to
 be 0.7 nm in water at room temperature) and  $\kappa\vx$ is the 
position-dependent screening length which depends on the total ionic concentration of the gel and salt ions, $\rho_{local}$. 
For a cross-linked nanogel network, where the monomer density is constant in space, $\rho_{gel}^c$ is given by  
Eq.~\ref{eq:rhogel} multiplied by the fraction of charged monomers $f_c$, whereas the salt charge density instead is again dictated by local charge neutrality, 
consistently with our previous choice of the Donnan potential to describe the electrostatic energy in the system.\\
%
%
Finally, the fourth term in the expansion of the free-energy functional depends on the excess 
free-energy density per particle $\epsilon^{exc}$, and measures the strength of protein-protein excluded-volume interactions~\cite{joe-langmuir}. 
In principle, the excess free-energy 
can be significant at moderate packing fractions and becomes very high close to the crystallisation
density of hard-spheres. However, these are well below the experimental packing fraction typically 
achieved in protein adsorption, at which $\epsilon^{exc}$ is  a relatively minor perturbation to the 
total free-energy with respect to all other terms present in the system 
(see for example Fig.~\ref{fig:initial}).
For this reason, we only consider its value in the second order expansion in density, 
the so-called $B_2$ approximation. Not only this further simplifies our calculations, 
but $B_2$ is also an experimentally measurable quantity which 
can be easily accessed from the osmotic pressure as a function of density for a protein solution.
Explicitly, this choice for $\epsilon^{exc}$ results in the following formula: 

\begin{align}
\FF^{exc} =& \sum\limits_p \int_V \epsilon( \{\rhox\} ) \rhox  \dx \nonumber \\
		 =& -{1\over 2 }\kb\sum_{i,j} B_2^{ij} \int_V \rho_i\vx \rho_j\vx \dx, 
\label{eq:B2expansion}
\end{align}

where the indices $i$ and $j$ run over all protein types in the system.
It was shown in \cite{joe-langmuir} that a reasonable value to take for $B_2$ is 
that for hard-spheres of the same mean size as the globular protein, given by:

\begin{equation}
B_2^{ij} = { 2 \pi \over 3 } ({\sigma_i + \sigma_j \over 2})^3
\label{eq:b2-definition}
\end{equation}

where $\sigma_i$ ($\sigma_j$) is the effective hard-core diameter of protein $i(j)$.
In principle, to account for polymer-protein excluded volume interactions, 
the sum in Eq.~\ref{eq:B2expansion} should include one term depending on the polymer 
density $\rho_{poly}$. The latter could also be considered another dynamic variable of the system, 
and its spatially dependent field treated at the same level as that of the protein, as 
done for example in \cite{fang-szleifer}.
Since for charged gels the polymer network is relatively rigid and the cross-linking distance is much larger 
than the protein size, we treat instead the polymer as a fixed effective excluded volume zone, 
and thus scale all protein densities $\rho_i$ in Eq.~\ref{eq:B2expansion} in the following way:

\begin{equation}
\rho_i(\vv{x}) \rightarrow  \xi\vx \rho_i(\vv{x}) = \left( \frac{1}{ 1- \rho_{poly}\vx v_{mon}} \right) \rho_i(\vv{x}),
\label{eq:scaled-rho}
\end{equation}

where $v_{mon}$ is the effective volume occupied by a monomer, which for our system is approximately 
$0.3$~nm$^3$ \cite{joe-langmuir}.
Outside of the gel, $\xi = 1$ and no scaling occurs, whereas inside the bulk polymer an increase 
in the number density of about $8\%$ is observed.


%
%
Finally, combining the previous definitions for the various terms appearing in Eq.~\ref{eq:Ftotal-decomposed} 
with Eq.~\ref{eq:chemPot}, we obtain for the chemical potential of the specie $p$ as a function of  $\rhoxt$:

\begin{align}
\bbb\muxt &=   \ln\left({\rhoxt \over \rho_{0} }\right) + \bbb\Delta G^{ads} S(\mid \vv{x}\mid) \nonumber \\
			& + \bbb V^{\mathrm{Don}}\vxt+\bbb V^{\mathrm{Born}}\vxt - \sum_j B_2^{pj} \rho_j\vxt. 
\label{eq:chemFinal}
\end{align}

By plugging Eq.~\ref{eq:chemFinal} into the generalised diffusion equation, Eq.~\ref{eq:diffusion}, we fully define the dynamics of our system,
which we will investigate later in Sec.~\ref{sec:numerical-results}.

\section{Diffusion timescales from simple analytical models \label{sec:compare}}

Before turning to fully solve the complex numerical equations described in the previous session, it is 
instructive to have at least a rough idea of the timescales involved in this problem by looking at a couple
of analytically solvable models. 

\subsection{Free diffusion in an open, spherically symmetric environment (Debye result) \label{sec:debye}}

When modelling adsorption phenomena, many authors resort to the famous 
Debye formula, which solves the problem of finding the steady-state
profile of a diffusing, non interacting specie around a spherically absorbing sink in 
contact with an infinite reservoir at density $\rho_p^{bulk}$. In practice, this require solving the following 
equation for the radial density of the specie $\rhor$:
%
%
\begin{equation}
\begin{cases}
 {1 \over r^2 } {\partial \over \partial r} r^2 {D_p \partial \rhor \over \partial r}  = 0 \\
\\
\rhor \mid_{r=R_{gel}} = 0 \\
 \\
\rhor \mid_{r=\infty} = \rho_p^{bulk}
\label{eq:diffusion-debye}
\end{cases}
\end{equation}
whose solution, assuming $D_p$ is constant in space, reads
\begin{equation}
\rho(r) =  \rho_p^{bulk} \left( 1 - {R_{gel} \over r } \right). 
\label{eq:diffusion-debye2}
\end{equation}

Given that this is a problem of simple diffusion with no terms apart the ideal one, 
the flux is equal to $J=-D_p{\partial \rho \over \partial x}$, from which follows the famous Debye 
formula for the steady-state flux:
\begin{align}
k_{debye} &= 4 \pi r^2 J(R_{gel}) \nonumber \\
 &= 4 \pi r^2 D_p {\partial \rho \over \partial r }\mid_{r=R_{gel}}  \nonumber \\
 &= 4 \pi R_{gel} D_p \rho_p^{bulk} 
\label{eq:flux-debye}
\end{align}
It should be emphasised that Eqs.~\ref{eq:diffusion-debye} and ~\ref{eq:flux-debye} 
describe adsorption by a perfectly adsorbing sink, whereby a particle, once it reaches the sink, disappears from the solution. 
Given that particles never accumulate at the boundary of the sink, and the bulk provide an infinite 
amount to replace those that are adsorbed, the flux is never zero and indeed these 
equations describe a non-equilibrium steady state problem. 
 
Whereas this formula can then approximate the flux for intermediate times (after a fast transient time
$t_{relax} = R_{gel}^2 / 2 D_p \approx 0.1$~ms for our system),  
in the real scenarios particles will accumulate at a boundary, generating a counter-gradient
that will in fact slow down and eventually stop diffusion. Hence care should be taken when estimating
protein loading speed using Eq.~\ref{eq:flux-debye}.
However, we note here that whereas mass conservation will slow down diffusion, other fluxes
present in the system not accounted in this simple description \textit{might} accelerate it, balancing the effect. 
Here we want to estimate the loading timescale for a specific case study: the adsorption of positively 
charged Lysozyme onto negatively charged PNIPAM nanogels. In this system, both electrostatic interaction and the
intrinsic adsorption energy speed up protein adsorption compared to ideal diffusion. 
Hence, in this particular case we expect a partial cancellation of errors to improve our estimate.

%
%
Given these premises, we will calculate as a measure of the speed of the loading kinetics the time taken by the 
nanoparticle to reach half the equilibrium loading, i.e. $t_{1/2}$.
To do this, however, we clearly require one important additional information, 
i.e. the total number of adsorbed particles at equilibrium. 
From experimental measurements \cite{joe-langmuir}, we know that 
about $5-7\cdot 10^4$ proteins are adsorbed on the nanogel. 
Since the number of adsorbed proteins per unit time (within this Debye approximation) 
is simply given by $N(t) = k_{Debye} t = 4 \pi R D_p \rho_p^{bulk} t$ 
we obtain by inverting this equation and setting $N = 6 \cdot10 ^4$, $D_p = 0.1~\mathrm{nm^2/ns}$, 
$R = 150$~nm and $\rho_p^{bulk}=2 \cdot 10^{-4}$~Mol a value of $ t_{1/2}\approx 1$~ms. 
As we will see, for an effect of cancellation of errors previously discussed, this
estimate will not be too far from the results obtained solving the much more complex DDFT equations.
%

\subsection{Free diffusion in a closed, spherically symmetric environment \label{sec:spherical-confinement}}
To account at least for mass-conservation effects within the bulk solution, 
we should solve the ideal diffusion equation under more realistic boundary conditions than those implied in the Debye treatment. 
Hence, we solve the diffusion equation for a closed, spherically symmetric environment.\\
We thus have, in spherical coordinates:

\begin{equation}
\begin{cases}
\prhort = {1 \over r^2 } {\partial \over \partial r} r^2 {D_p \partial \rhort \over \partial r}  \\
\\
\prhort \mid_{r=R_{core}} = 0 \\
\\
\prhort \mid_{r=L} = 0 \\
\\
\rhort \mid_{t=0} = \rho_p^{bulk} \theta[r-R_{gel}]
\end{cases}
\label{eq:diffusion-simple}
\end{equation}

where $R_{core}$ is the radius of the nano particle hard-core (see Fig.~\ref{fig:system-schematic}), 
and the outer boundary $L$ depends on the nanogel number density $\rho_{np}$, 
as specified later in Sec.~\ref{sec:numerical-results}.
The initial density profile is taken to be a homogeneous density equal 
to the initial bulk density value $\rho_p^{bulk}$, except in the nanogel where it is taken to be zero, corresponding 
to a possible setup where nanoparticles are inserted in an otherwise equilibrated solution of proteins. 
This problem can be fully solved analytically by standard Fourier techniques. 
We will only report here the final form of the solution for clarity, where we also assumed $D_p$ to be
constant in space

\begin{align}
&\rhort = C_0 + {1\over r}\sum_{n=1}^{n=\infty} \exp{\left( -\lambda_n^2 D_p t \right)} N_n^{-1} C_n \phi_n(r) \\
&N^{-1}_{\lambda_n} = \\
&{2 \over \left[ \left ( \lambda_n^2 + {1\over R_{core}^2 }\right) \left( \left( L - R_{core} \right) - {1 \over L \left ( \lambda_n^2 + {1\over L^2 }\right) }\right) \right] + {1\over R_{core}}} \nonumber \\
&C_n = \int_{R_C}^{L} r' \theta \left(r'-R_{gel}\right) \phi_n(r') dr' \\
&\phi_n(x) = \lambda_n \cos\left(\lambda_n r\right) + {1\over R_{core}} \sin\left(\lambda_n r\right)
\label{eq:diffusion-spherical}
\end{align}

where $C_0$ is nothing but the average value of the initial density in the domain, i.e.

\begin{align}
C_0 = {3 \over 4 \pi \left( L^3-R_{core}^3 \right) } \int_{R_{core}}^L 4 \pi r^2 \rho_p^{bulk} \theta \left(r-R_{gel}\right) dr,
\label{eq:C_0}
\end{align}

and $\lambda_n$ is given by the solution of the following transcendental equation

\begin{align}
\tan\left( \lambda_n \left(L-R_{core}\right) \right) = { \lambda_n \left(L-R_{core}\right) \over 1+ \lambda_n^2 L R_{core}}, 
\label{eq:lambda}
\end{align}

where $n$ labels the infinitely many solution for this equation.\\*
The solution to this problem is quite instructive, and we discuss some of its main features here.
First of all, a timescale $\tau_D = { \left(L-R_{core} \right)^2 \over D_p }$ appears. 
Note that this timescale does not contain any reference to $R_{Gel}$, i.e. the radius of the nanogel. 
Moreover, at the typical densities encountered in experiments, one has that $L>>R_{core}$, 
hence the only relevant timescale is controlled by the nanogel average distance $L$, itself a function of the nanogel density, 
$L \approx \rho_{np}^{-1/3}$ (see Sec.~\ref{sec:numerical-results}). 
This would mean that the adsorption kinetics for micron- or nano-sized gels, if measured at the same number density, 
will be the same within this model. If experiments instead are made at constant packing fraction $\rho V_{nanogel}$, 
which scales as $R_{gel}^3$, than the loading dynamics will be many orders of 
magnitude faster for nanogels. This can partially rationalise the very different timescales 
observed in the experiments for these two systems \cite{nicole-kinetics,lysozime-soft-matter}.
If we plug into the definition of $\tau_D$ the values of $L$ for the experiments we are trying to describe \cite{joe-langmuir} (see Sec.~\ref{sec:numerical-results}), 
which is about $10^3$~nm, and the diffusion coefficient of lysozime in water, which is of order $D_p\approx 0.1 \mathrm{nm}^2/\mathrm{ns} $
\cite{lysozime-soft-matter}, by truncating Eq.~\ref{eq:diffusion-spherical} to the first few terms in $n$, we obtain an estimate of 
$t_{1/2} \approx  2\cdot 10^{-3} \tau_D = 2 \cdot 10^{-2}$~ms. \\
The reason for which diffusion is here much faster than for the Debye case is that we properly took into account the 
full density evolution, which has initially a strong density gradient -hence associated flux- at the nanogel/solution boundary, 
whereas in the Debye case we simply used the steady state value of the flux to calculate the loading. 
Regardless, we will see later in Sec.~\ref{sec:numerical-results} how neither the timescales nor 
the density profile obtained from the solution of the ideal diffusion equation correspond to what is observed for 
our DDFT model, warranting that ideal diffusion equations should be taken very carefully when used as an interpretative model for 
experimental data, even from a qualitative point of view. 

\section{Numerical results from the DDFT equations \label{sec:numerical-results}}

In this section we will present a series of results from the full numerical  solution of the DDFT equation.
The associated PDE for the time-evolution of the density field was solved by discretising the problem on a fixed grid of spacing
$0.5$~nm and propagating the equation of motion using a 4th order Runge-Kutta method with a timestep in the range $[0.025-0.05]$~ns
depending on the parameters. Simulations were run for a number of timesteps in the range $[10^7-10^9]$, 
and for all of them mass was conserved within less than a $1\%$ error.\\
The boundary conditions to solve Eq.~\ref{eq:diffusion} are dictated by our system. 
One of the boundaries is the nanoparticle hard-core on which the polymeric
gel is grafted. For all intense and purposes, this core can be safely regarded as a barrier 
that proteins cannot penetrate. A no-flux boundary condition at $r=\Rg$ takes care of that.
The second boundary is given by the experimental setup we want to describe.
In a real experiment, nanoparticles are found in solution at a low but finite density,
and in principle their exact position will matter for the protein adsorption dynamics: the full problem 
would couple the position of all nanoparticles to the protein density field. Instead of solving this 
very complex computational problem, we take a statistical approach and use instead a cell-model 
\cite{joe-langmuir}. Each nanoparticles is supposed to be isolated in a spherical cell of fixed volume and 
the sum of all volumes must fill the whole space, giving the following condition for the cell 
radius $R_{cell}$:

\begin{align}
N_{np} V_{cell} &= V_{tot} \nonumber \\
&\rightarrow R_{cell} = \left( 3 V_{tot} \over 4 \pi N \right )^{1/3} = \left( 3 \over 4 \pi \rho_{np} \right )^{1/3}
\label{eq:radius-domain}
\end{align}

where $N_{np}$ is the number of nanoparticles present in solution and $\rho_{np}$ 
their number density. This is a valid assumption when nanogels do not tend to aggregate but remain dispersed.
In this model, a no-flux boundary condition naturally arises at $r=R_{cell}$ , because the radial flux from 
neighbouring cells exactly compensates.\\*

To allow for the tightest possible comparison to experiments, we will analyse the same 
system as in Ref.~\cite{joe-langmuir}. 
Briefly, a nanogel with a hard-core radius of $R_{core}\approx 60$~nm with a charged polymer corona of $90$~nm, 
hence $\Rg \approx 150$~nm. There are approximately $3.7 \cdot 10^6$ monomers for each nanogel, about which 
$4.9 \cdot 10^5$ carry a net charge of $-1e$ (i.e. $f_c\approx 13\%)$, for a total charge density of $\rho_{gel}^c\approx 4  \cdot 10^{-2}~e / \mathrm{nm}^3$. 
For comparison, the average concentration of cations (anions) due to the dissociated salt is almost an order of 
magnitude smaller, i.e. $\rho_{s} = 7~\mathrm{mMol}$, or $\approx 4 \cdot 10^{-3}~e/ \mathrm{nm}^3$. The volume of each monomer 
is estimated to be about $0.3~\mathrm{nm}^3$ so that the total excluded volume in the gel is $\approx 8 \%$. 
The number concentration of nanogels is $\rho_{np} = 8.42  \cdot10^{-10}~\mathrm{M}$, i.e. about $1/\mu \mathrm{m}^3$. 
This concentration is related to the average distance between gel particles by Eq.~\ref{eq:radius-domain}, 
which gives $R_{cell}\approx 780$~nm, about 5 times the radius of the gel itself.
When not specified otherwise, the protein under investigation is Lysozyme, which carries a net charge 
of $+7 e$ at  the pH$= 7.2$ considered. The initial bulk number density of protein is taken to be 
$\approx 5 \cdot10^4 \rho_{np}$, corresponding to $5 \cdot10^{-5} /$ nm$^3$.
The diffusion constant of Lysozyme in water is taken to be $10^{-1}\mathrm{nm^2/ns}$, in accordance with 
both experimental and theoretical values in the literature \cite{lysozime-diff,lysozime-diff2}, whereas
that in the gel it is taken to be an order of magnitude slower, a reduction consistent to that 
observed in other similar polymeric systems \cite{lysozime-soft-matter}. The only additional necessary parameter to model the 
kinetics is the intrinsic adsorption energy $\Delta G^{ads}$ of the protein, 
which can be extracted from experiments probing the thermodynamics of protein adsorption for the same system \cite{joe-langmuir}. 
For Lysozyme, this was determined to be equal to $7.25 k_{\mathrm{B}} T$.

Before we proceed to discuss the results of our numerical modelling, we should point out that in order to simplify the problem 
our model does not take into account the fact that the polymer gel can shrink upon protein adsorption. Experimentally, for the 
initial bulk protein concentration studied in this case, the maximum reduction in the polymer radius, achieved at equilibrium, is roughly $10\%$ \cite{joe-langmuir}. 
Since the polymer volume, and hence the number of protein's adsorption sites (in the sense specified in Ref.~\cite{joe-langmuir}), 
turns out to be an important quantity to get a realistic estimate of the loading kinetics, we take as the fixed value for the radius of the gel 
the equilibrium value. Whereas this simplification might change the exact numerical results, it does not impact in a significative
way our estimates for the orders of magnitude nor the trends observed.
\subsection{Equilibrium \label{sec:equilibrium}}
\begin{figure*}[h!]
  \centering
  \includegraphics[width=0.48\textwidth]{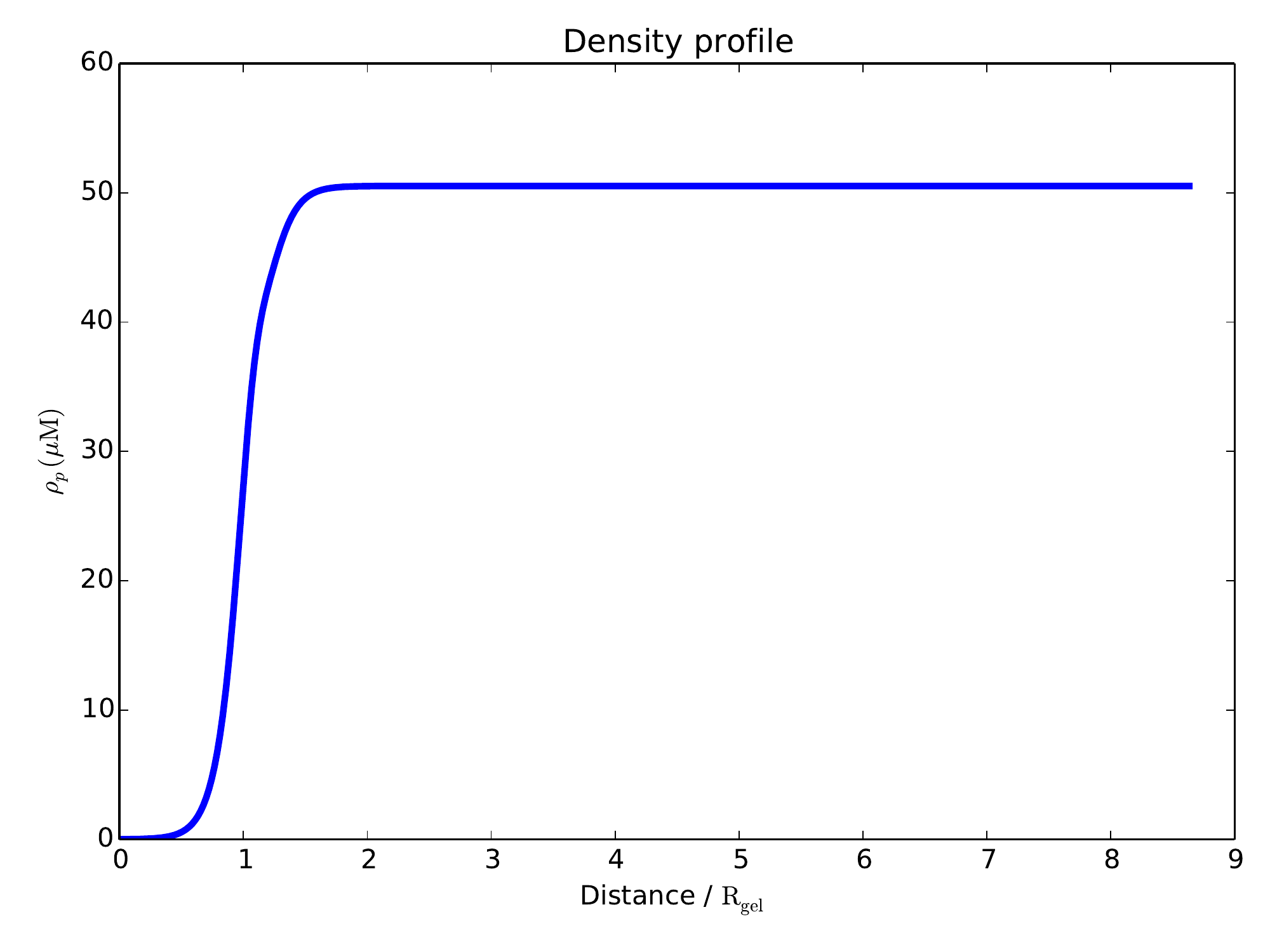} \includegraphics[width=0.48\textwidth]{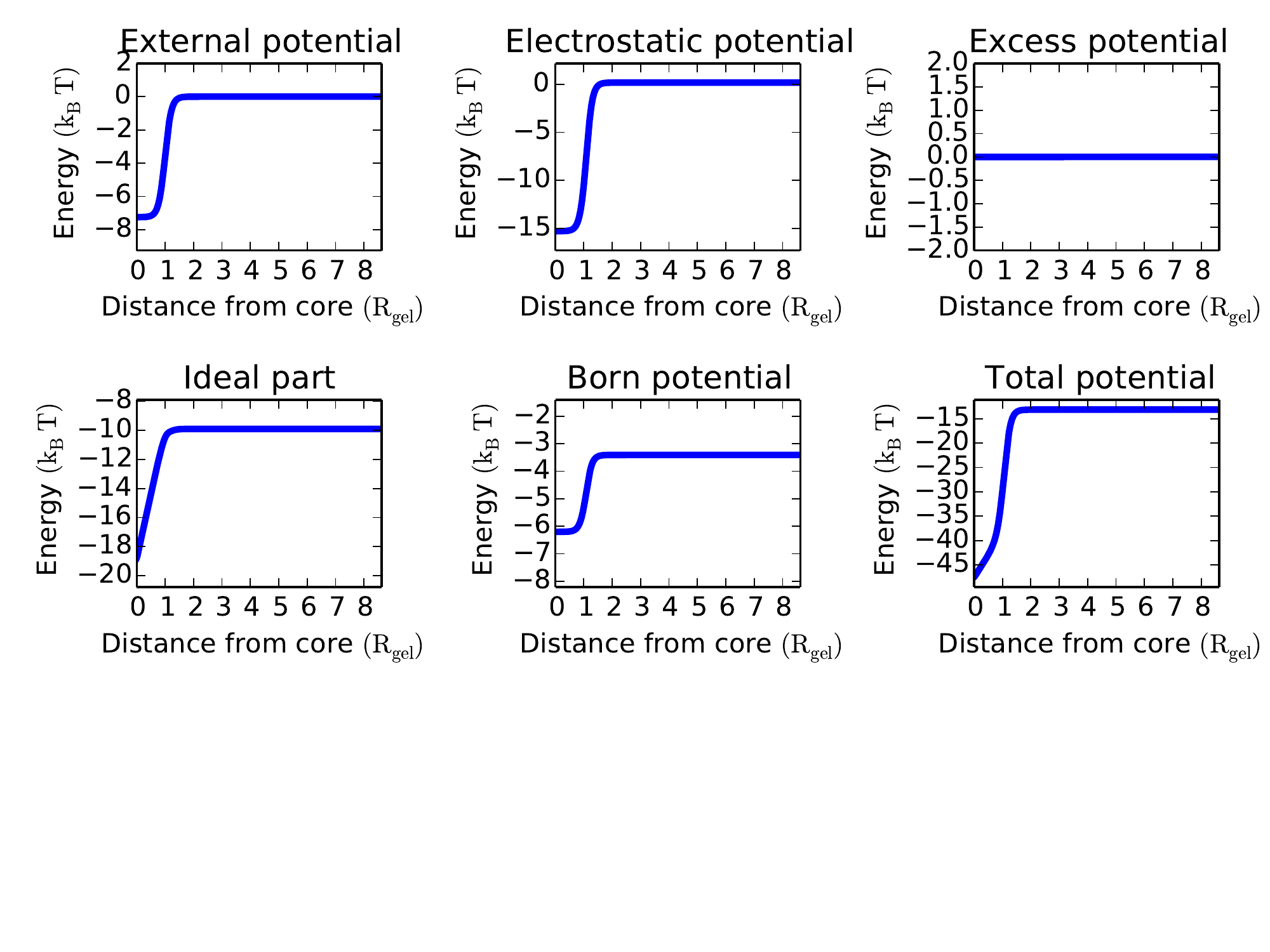}\label{subfig:a}\\
  \includegraphics[width=0.48\textwidth]{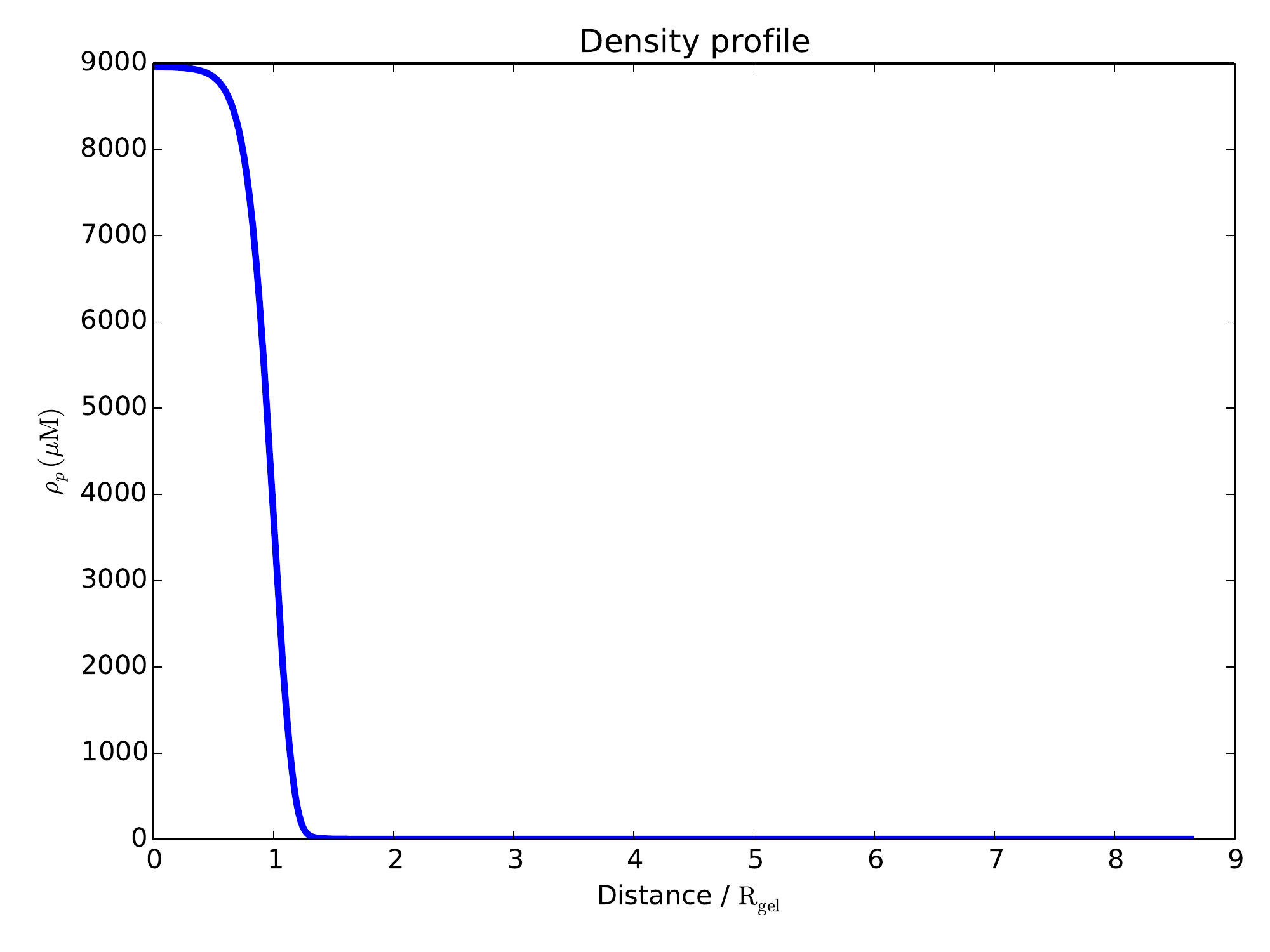} \includegraphics[width=0.48\textwidth]{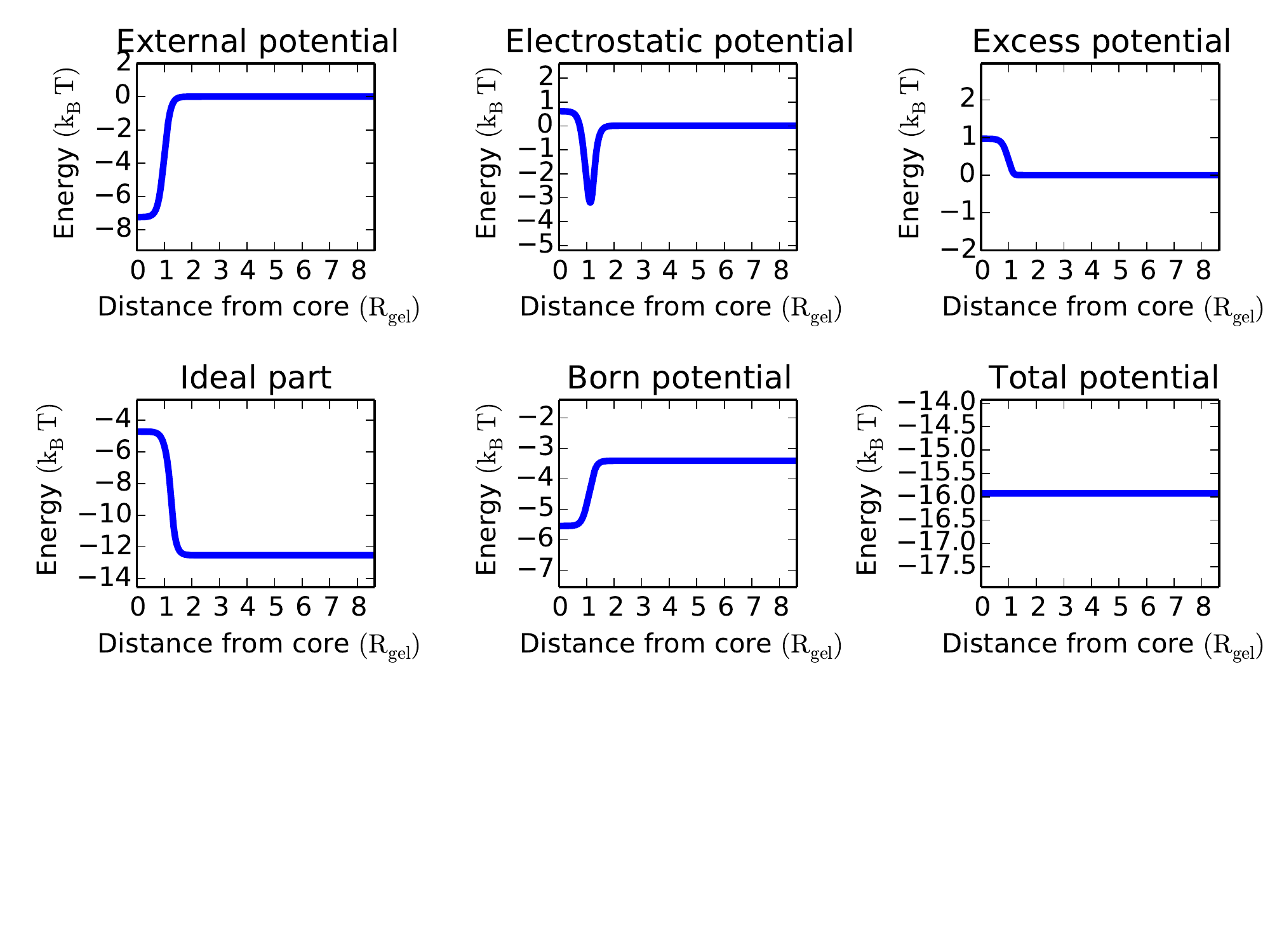}\label{subfig:b}
\caption{Density profile and spatial variation of the different terms in the chemical potential, 
Eq.~\ref{eq:Ftotal-decomposed}, for the initial (top) and equilibrium (bottom) density. Note 
that both the ideal chemical potential and the electrostatic term change 
slope at the gel/bulk boundary (R=1), hence the associated thermodynamic 
force must change from attractive to repulsive in the course of the simulation. 
Such a dynamical change in the forces driving adsorption cannot be 
captured with simpler models based on Langmuir kinetics, since
in the latter all effects are gathered in a single equilibrium constant \cite{fang-szleifer-compare}.
}
\label{fig:initial}
\end{figure*}

Before discussing the dynamics of our system, it is of interest to 
look at the final equilibrium solution, in order to highlight the role 
played by the various term in determining the final equilibrium density profile \rhoxe.
This can be obtained by looking for the density profile for which $J(\vv{x}) = 0$, or, equivalently,
minimising the free-energy functional (Eq.~\ref{eq:Ftotal}) under the constraint of a fixed number of proteins, leading to:

\begin{align}
&\rhoxe  = \nonumber \\
&\rho_0 e^{\left( -\beta\mu^{eq}_p \right)} e^{\left[- \beta \left(V^{ext}(\vv{x}) + \epsilon^{exc}\left(\{\rho^{e}_{p}(\vv{x})\}\right) + \rhoxe {\partial \epsilon^{exc}\left(\{\rho^{e}_{p}(\vv{x})\}\right) \over\partial \rho_p} \right) \right] }.
\label{eq:equirho}
\end{align}

where the quantity $\rho_0 e^{\left( -\beta\mu^{eq}_p \right)}$ is determined by
imposing a fixed number of proteins $N_p$ in our cell volume for each species in the system, i.e.

\begin{equation}
\int_{R_{core}}^{R_{cell}} 4 \pi r^2 \rhore  dr = N_p = V_{cell} \rho_p^{bulk}.
\label{eq:mass-conservation}
\end{equation}

where $\rho_p^{bulk}$ is the \textit{initial} bulk concentration of protein $p$ (note that due to mass 
conservation, the density of proteins in the bulk will diminish due to adsorption onto the nanoparticle).
In general, when inter-particle interactions are present and hence the density appears 
in both sides of Eq.~\ref{eq:equirho}, a closed formula for $\rho^{e}_p$ cannot be found, 
and the problem must be solved iteratively starting with a trial density and iterating until 
self-consistency is achieved.\\
We report in Fig.~\ref{fig:initial} 
the value of the various terms in Eq~\ref{eq:Ftotal-decomposed} for the 
initial (top) and equilibrium (bottom) density distribution.
%


An important feature to notice in these profiles is the change of some of the thermodynamic
forces in their slope at the gel/bulk solution boundary (i.e. R=1), since this is 
related to the adsorption flux through the equation
\begin{equation}
J_p^{i}(t) = -D_p(R_{gel}) \rho_p(R_{gel},t) {\partial \beta \mu_p^i(R_{gel},t) \over \partial r}
\label{eq:flux-force}
\end{equation}
where the superscript $i$ labels the specific thermodynamic potential considered (e.g $V^{\mathrm{Don}}, V^{\mathrm{Born}},...$). 
Indeed, this change means that whereas initially all thermodynamic forces drive the system
towards absorbing protein, closer to equilibrium only the Born and intrinsic energy term favour
adsorption, whereas the ideal and excess terms, as well as the Donnan potential, prevent it.
It is the balance between these opposing terms that determines the final equilibrium, and strongly 
influences the observed dynamics of the system.\\*
\subsection{ Dynamical behaviour \label{sec:loading-vs-exp} }
\begin{figure*}[h!]
\includegraphics[width=0.48\textwidth]{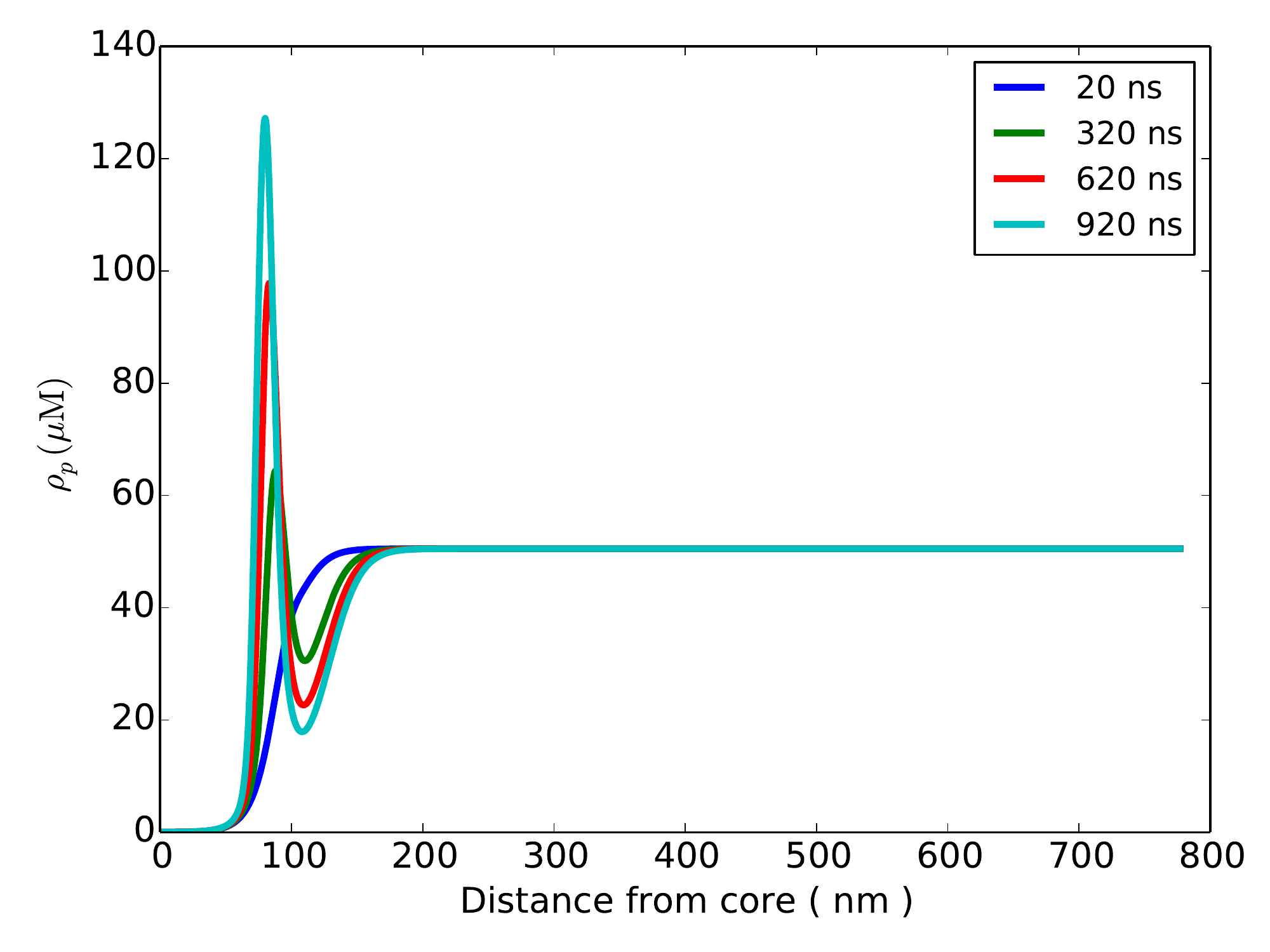}\label{fig:lyso-short}\\
\includegraphics[width=0.48\textwidth]{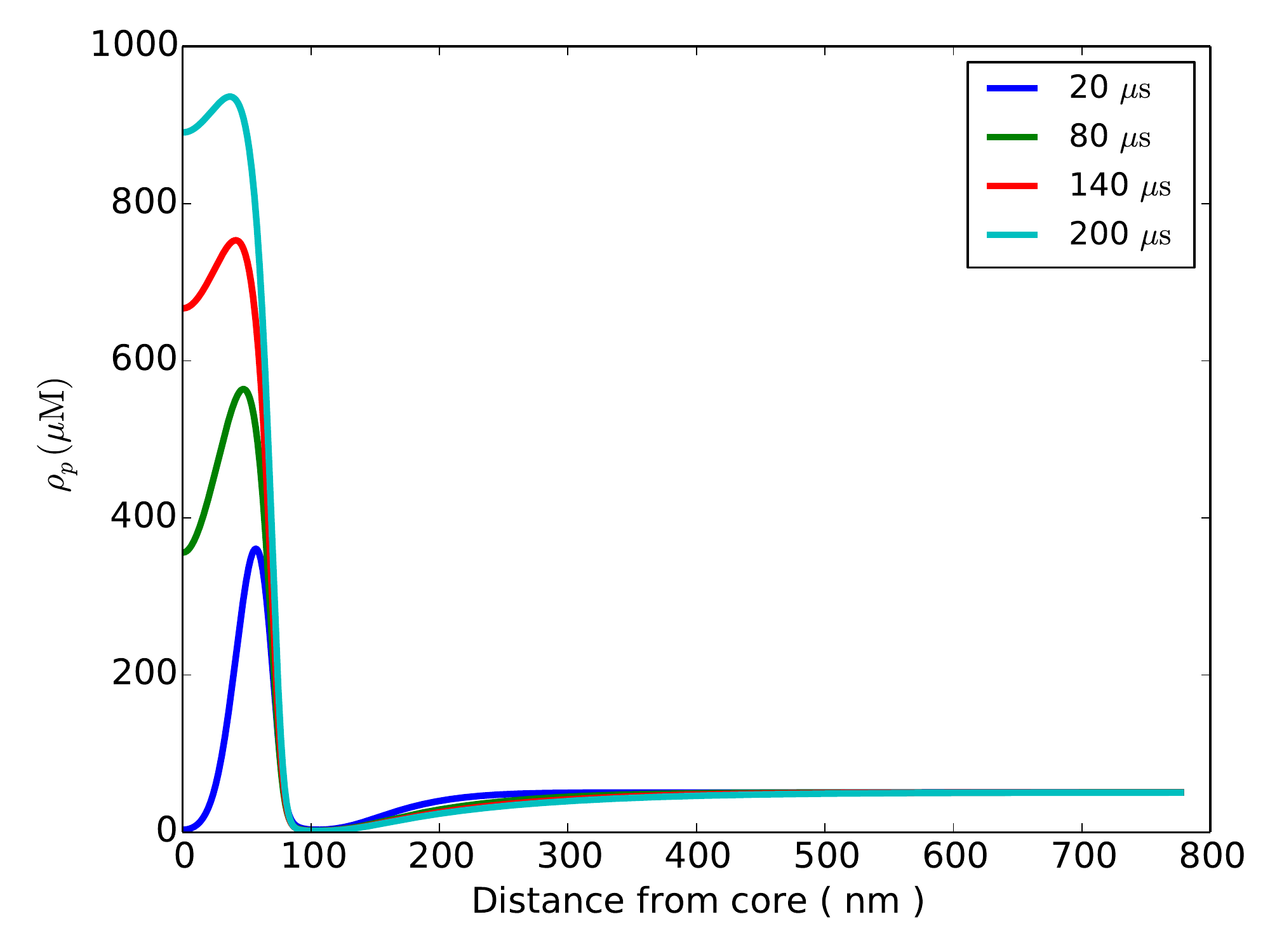}\label{fig:lyso-medium}\\
\includegraphics[width=0.48\textwidth]{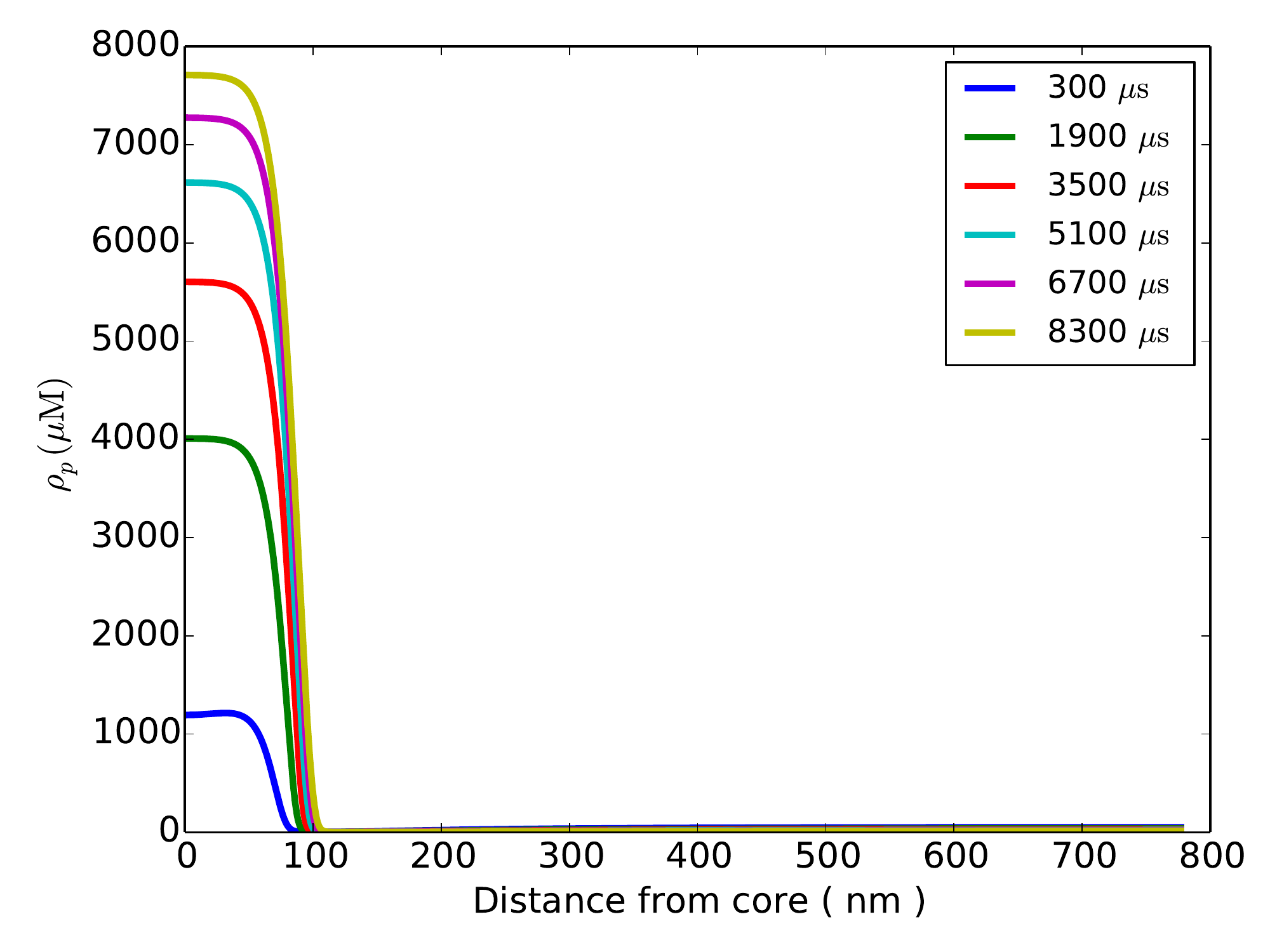}\label{fig:lyso-long}
\caption{Evolution of the density profile at short, intermediate and long timescales (see explanation in text).}
\label{fig:ddft-diffusion}
\end{figure*}

We report the full time evolution of the density profile for the system in Fig.~\ref{fig:ddft-diffusion}.
Let us first discuss these profile qualitatively.
Three distinct regimes can be observed. At very short timescales ($t<10\mu$s), a density instability is 
generated at the boundary between the gel and bulk surface, which propagates towards the nanoparticle hard core. 
This density peak stems from competition between a very strong energy gradient at the gel-bulk boundary 
which pushes protein towards the gel together with the reduced diffusion coefficient in the gel region, 
which is about $1/10$ that in the bulk solution, which causes proteins to accumulate at the interface. 
At intermediate timescales ($t<110\mu$s), the density peak diffuses far enough towards the gel/hardcore boundary, 
an appreciable concentration of protein builds up in this region, and the density peak becomes more diffuse, 
eventually reaching a width approximately equal to the gel width. At this point, a step-like density profile is obtained, 
and at later times the only qualitative change in the density profile is its height, which grows in time until the full equilibrium loading is reached.\\*
A question that naturally arises is whether a similar dynamical behaviour can be reproduced using a simple diffusion 
model where only the ideal term is retained, but we still account for the space-dependence of the diffusion 
coefficient to make a fair comparison. This is what many kinetic models of protein adsorption assume 
either implicitly or explicitly, completely neglecting the role of energy gradients in the system \cite{lysozime-soft-matter}. 
Fig.~\ref{fig:simple-diffusion} reports for comparison the evolution of the density profile for the same type of 
protein described in Fig.~\ref{fig:ddft-diffusion} but described in terms of the ideal diffusion equation,
where the diffusion coefficient has been taken to have the same spatial dependency as for the DDFT model
to allow for proper comparison:

\begin{figure*}[h!]
\includegraphics[width=0.48\textwidth]{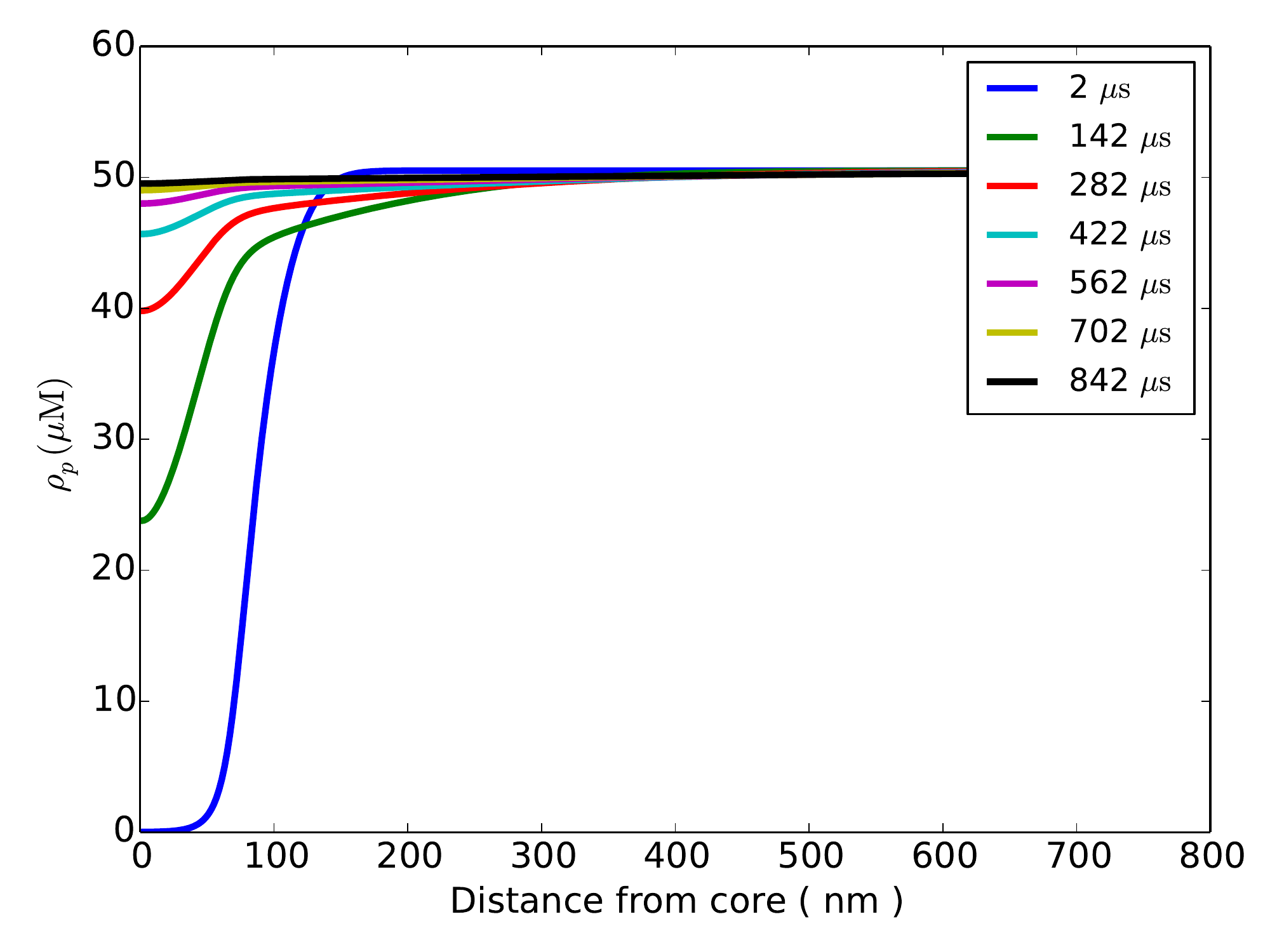}
\caption{Density evolution for an ideal model with spatially dependent diffusion coefficient.}
\label{fig:simple-diffusion}
\end{figure*}

\begin{figure*}[h!]
\includegraphics[width=0.48\textwidth]{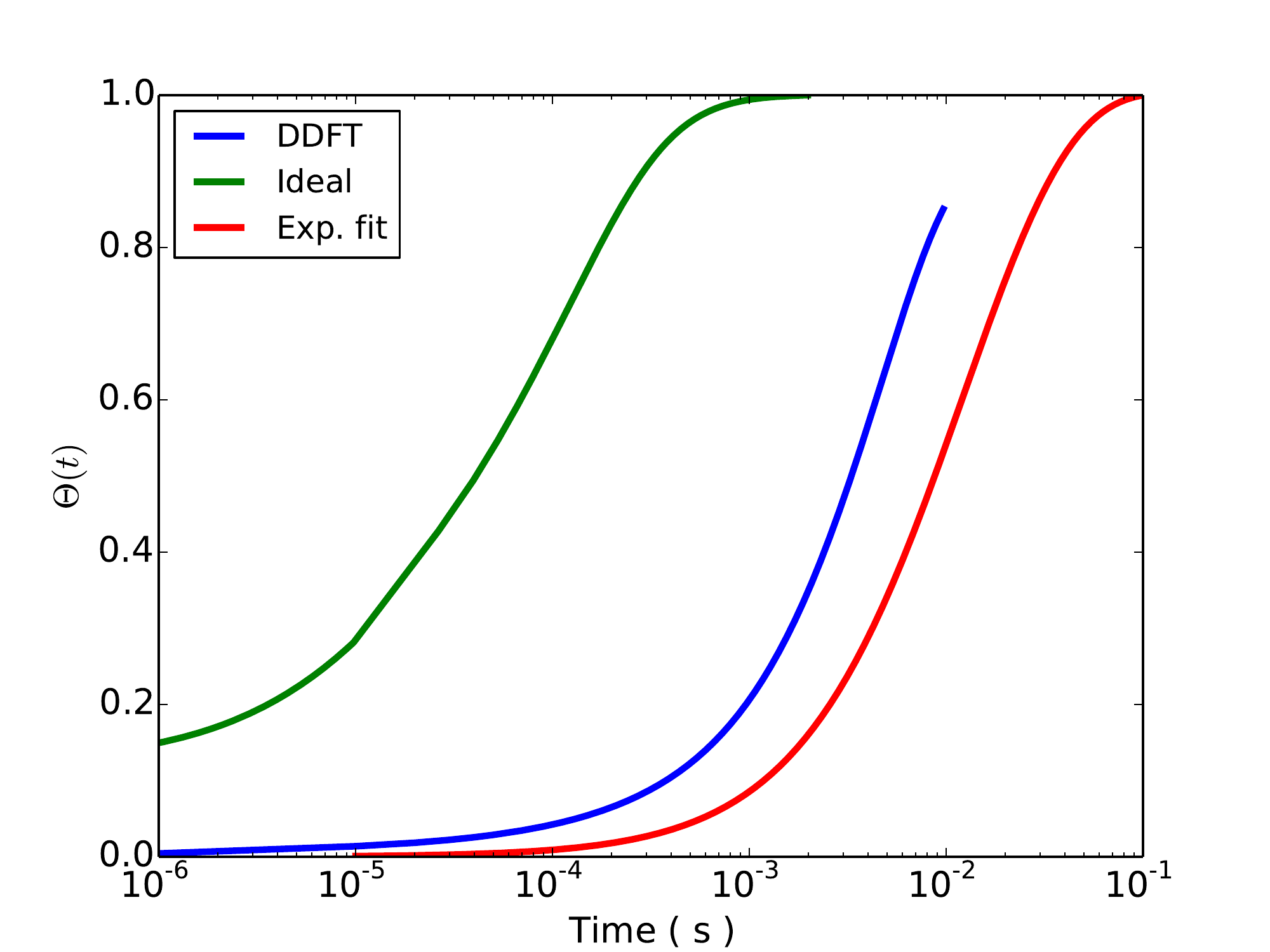}
\caption{Comparison of $\Theta(t)=N(t) / N(\infty)$ vs. time $t$ for a model based on ideal diffusion, our DDFT model and 
for a Langmuir model fit to reproduce the experimental data in \cite{nicole-kinetics}.}
\label{fig:load-compare-exp}
\end{figure*}

It is evident that the dynamics is not only just quantitatively approximate but also qualitatively 
very different compared to the one obtained using a more realistic model. 
Moreover, the timescales are clearly off by more than an order of magnitude, given 
that the density profile for purely ideal diffusion has almost reached its equilibrium value 
in half a millisecond, in contrary to the full description where at five milliseconds the density 
profile is still relatively far from being equilibrated. 
In principle, one could argue that DDFT models might not reproduce the loading dynamics 
better than the ideal diffusion equation. To show this is not the case, we report for both models in Fig.~\ref{fig:load-compare-exp}
the time dependence of the loading $\Theta(t) = N(t) / N(\infty) $ (where $N(t)$ is the number of adsorbed proteins, obtained by simply  integrating the density 
over the whole gel volume) and compare it to that extrapolated from fitting of experimental data, as
shown in Ref.~\cite{nicole-kinetics}. In this latter paper, it was shown that an empirical Langmuir fit was able to
reproduce, using the same parameters, data at different densities. 
In order to compare our data with those from experiments, we scaled the experimental value to the same 
protein density studied here \footnote{simulations of the density for which experimental data is directly available is not 
possible since this would require simulating timescales a couple of orders of magnitude higher than those 
accessible within our model, due to computational limitations}.

It should be clear from Fig.~\ref{fig:load-compare-exp} that our DDFT
description, although still not in complete quantitative agreement with experimental data, is a much better representation 
then an ideal diffusion model, where the dynamics is off by more than one order of magnitude.

\begin{figure*}[h!]
\includegraphics[width=0.48\textwidth]{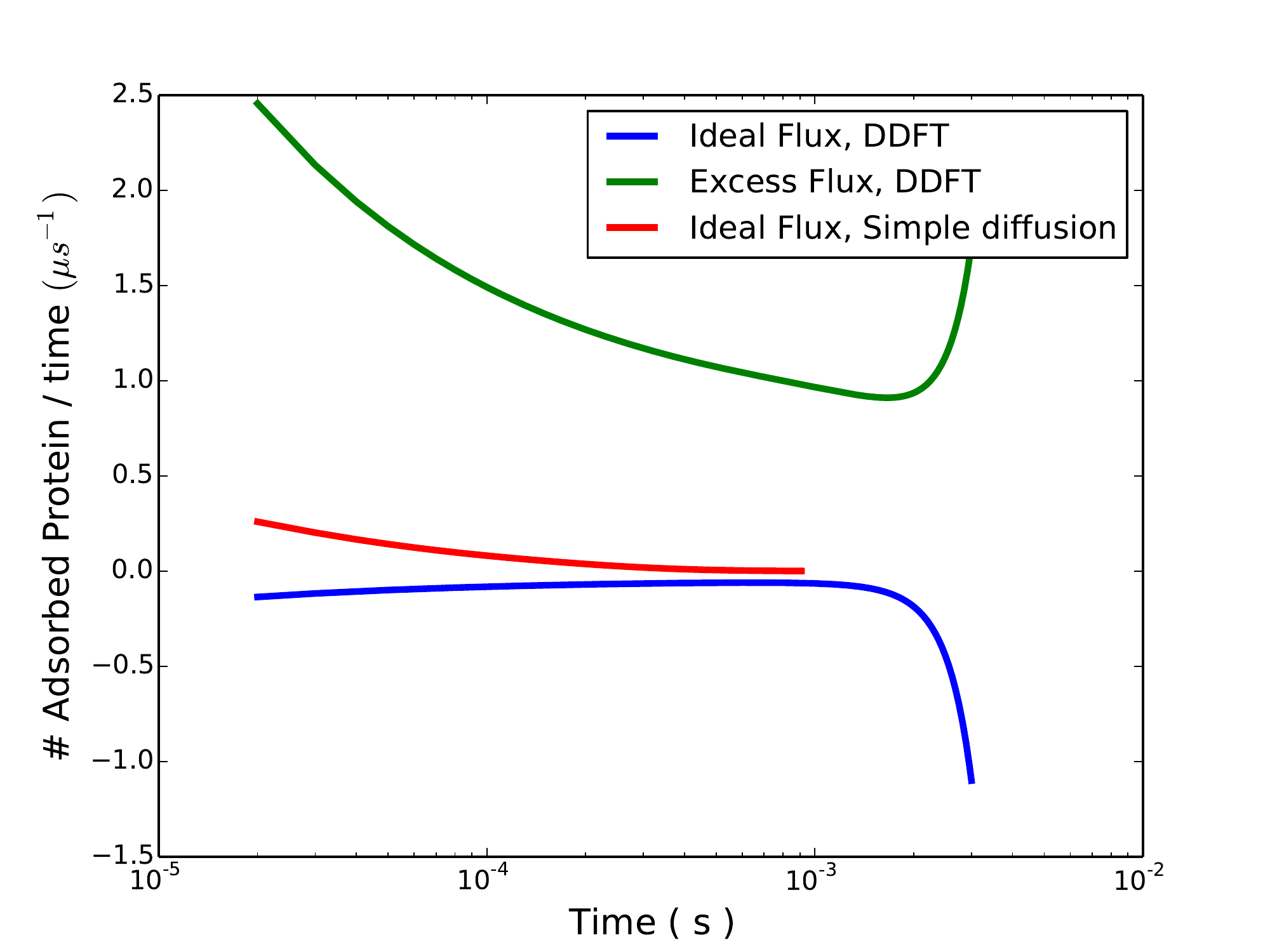}\\
\includegraphics[width=0.48\textwidth]{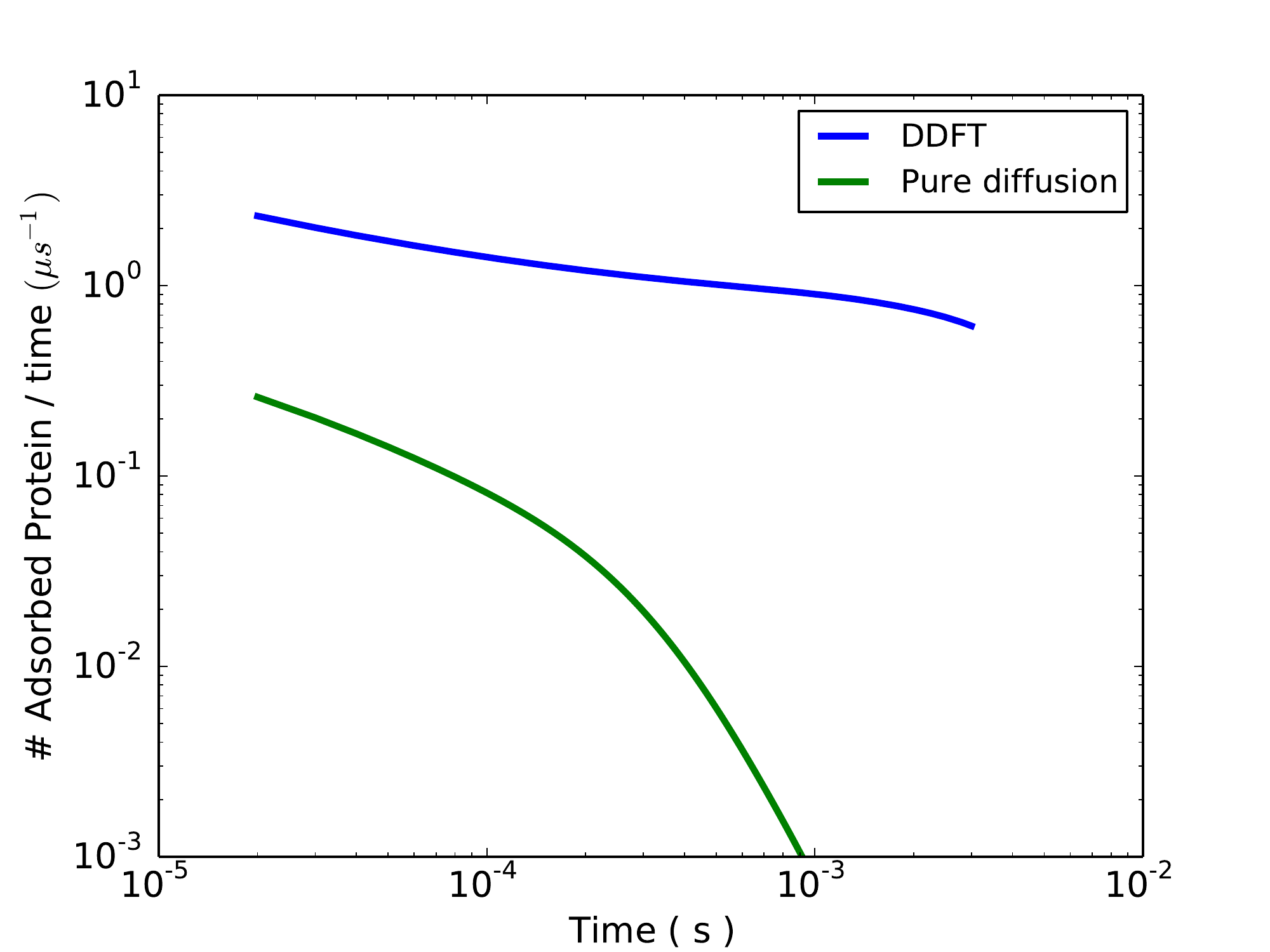}
\caption{Fluxes as a function of time. Top) Ideal and excess flux in our DDFT model vs ideal flux in the model of Fig.~\ref{fig:simple-diffusion}. 
Bottom) Total flux in our DDFT model and in the purely diffusive case.}
\label{fig:compare-fluxes}
\end{figure*}

Since an important fact is that  ideal diffusion completely neglects the important 
fluxes due to energy gradient in the systems, it is illuminating to look at how much these contribute to 
protein loading, as shown in Fig.~~\ref{fig:compare-fluxes},
where we plot the ideal and excess protein flux at the gel-solution boundary 
(the sum of which, by integration over time, gives the loading). 
As observed in Fig.~\ref{fig:compare-fluxes}, the ideal flux in both models are similar. However, 
the real flux is the sum of the ideal \textit{and} excess flux, 
the latter being zero in an ideal diffusion model. In this regard, we notice how the excess flux 
is always at least comparable if not dominant w.r.t the ideal one, with the result that 
not taking it into account leads to a wrong estimate of the loading.\\
In our model, the excess flux is always positive, hence it leads to a higher number of adsorbed proteins
per unit time in the DDFT scenario. This is not in contrast with ideal diffusion models relaxing to equilibrium 
much faster than the more realistic DDFT description because the equilibrium number of 
proteins calculated within an ideal model is orders of magnitudes smaller than that from DFT.
In fact, underestimation of the equilibrium amount of protein is possibly the largest source 
of error in using the ideal diffusion equation to model protein adsorption \cite{lysozime-soft-matter}, since it can only predict
a final flat equilibrium profile where the density is constant throughout the system. However, as expected 
from simple thermodynamics arguments, a non-homogeneous density must appear 
whenever any type of gel/protein interaction is present.
Hence, care should be taken when using ideal diffusion models to analyse experimental data.
For example, in Ref.~\cite{lysozime-soft-matter} Li \textit{et al.} found that in order to obtain 
the correct timescales, they had to assume the presence of trapping binding sites that reduce the
mobility of the proteins, effectively inducing a diffusion constant about 2 to 3 orders of magnitudes lower
than that expected for similar polymer/protein systems. 
Such a small value is probably an artefact arising from not 
including any electrostatic driving force in their description, since in our DDFT model we were able to obtain
the correct timescale without assuming such a surprisingly small diffusion coefficient. 
The importance of electrostatics is pointed out by the fact that, in the same experiments, 
they found that the number of expected binding sites (which determines the effective diffusion 
coefficient) is strongly dependent on pH, varying by a factor of 20 in the pH range $[3-7]$~\cite{lysozime-soft-matter}.\\

\subsection{ Parametric study \label{sec:parametric} }
Given that we observe both qualitative and semi-quantitative agreement with experiments, 
we can confidently use the current model to investigate the sorption kinetics for different scenarios,
and try to rationalise the observed trends. 
In particular, we assess here how the dynamics changes as a function of four important  
parameters characterising our system, i.e. protein valence, nanoparticles and protein's concentration 
and intrinsic adsorption energy.
We do this by looking at both the unnormalised and normalised amount of adsorbed proteins, $N(t)$ and $\Theta(t)$. 
As previously done in Sec.~\ref{sec:compare} we will take as an informative quantity 
to measure the speed of the kinetics the time to achieve half the equilibrium loading, $t_{1/2}$, 
which we report in Table \ref{tab:time}.
 The standard values for the parameters in the following simulations are $\beta\Delta G^{ads} = -1$, $\rho_p = 2.02\cdot
 10^{-4}$~M, $\rho_{np}=3.36 \cdot 10^{-9}$~M and $Z=1$, and in each set of simulations one of this quantity is
 varied keeping the other fixed. The parameters describing the nanoparticle, such as its radius or that of the polymer
 gel coating it, are the same as those for the Lysozyme model.
Fig.~\ref{fig:compare-all},\ref{fig:compare-all-unnorm} and Table~\ref{tab:time} summarise our results:

\begin{figure*}[h]
\includegraphics[width=0.48\textwidth]{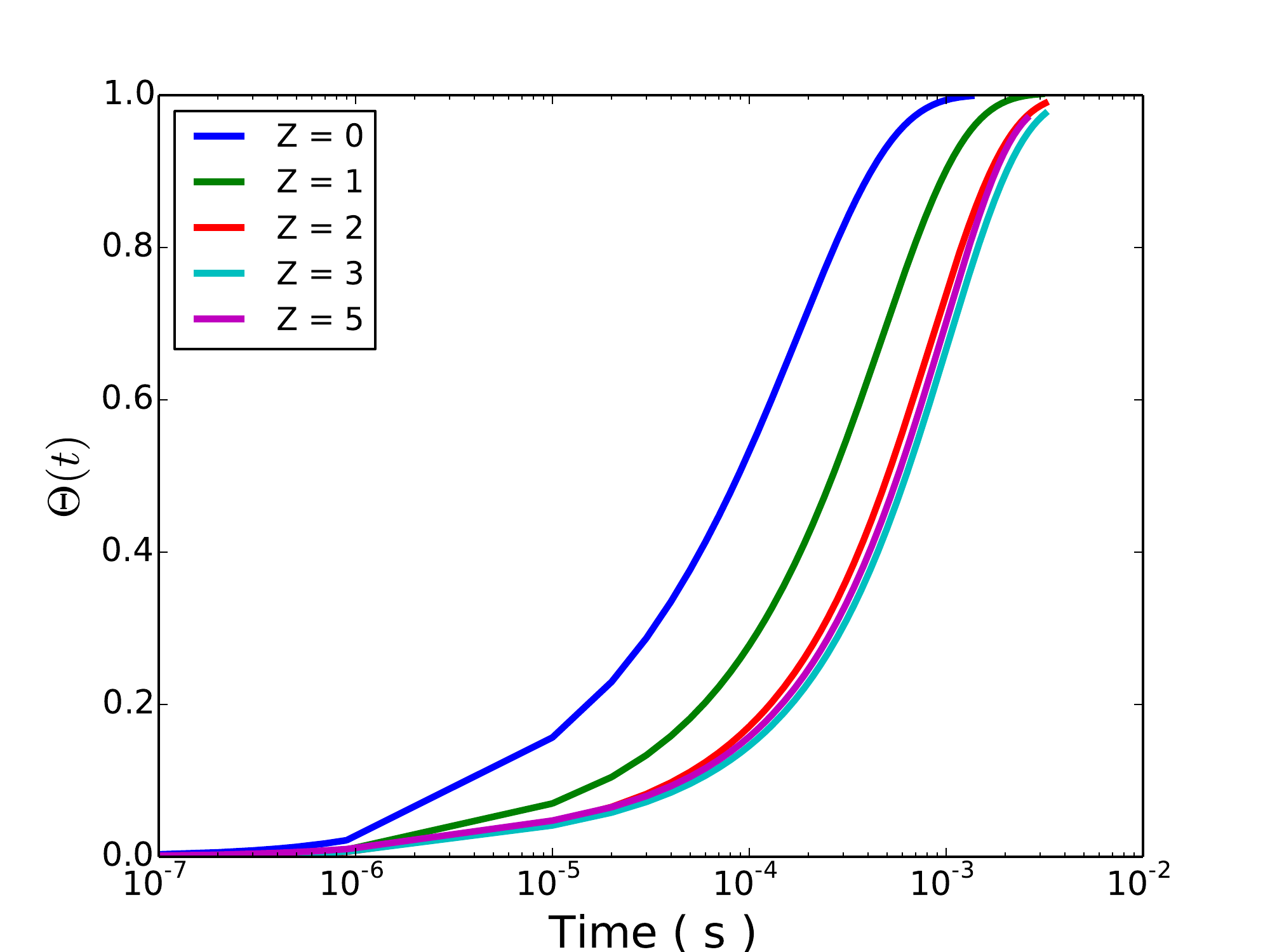}\label{fig:valence1}
\includegraphics[width=0.48\textwidth]{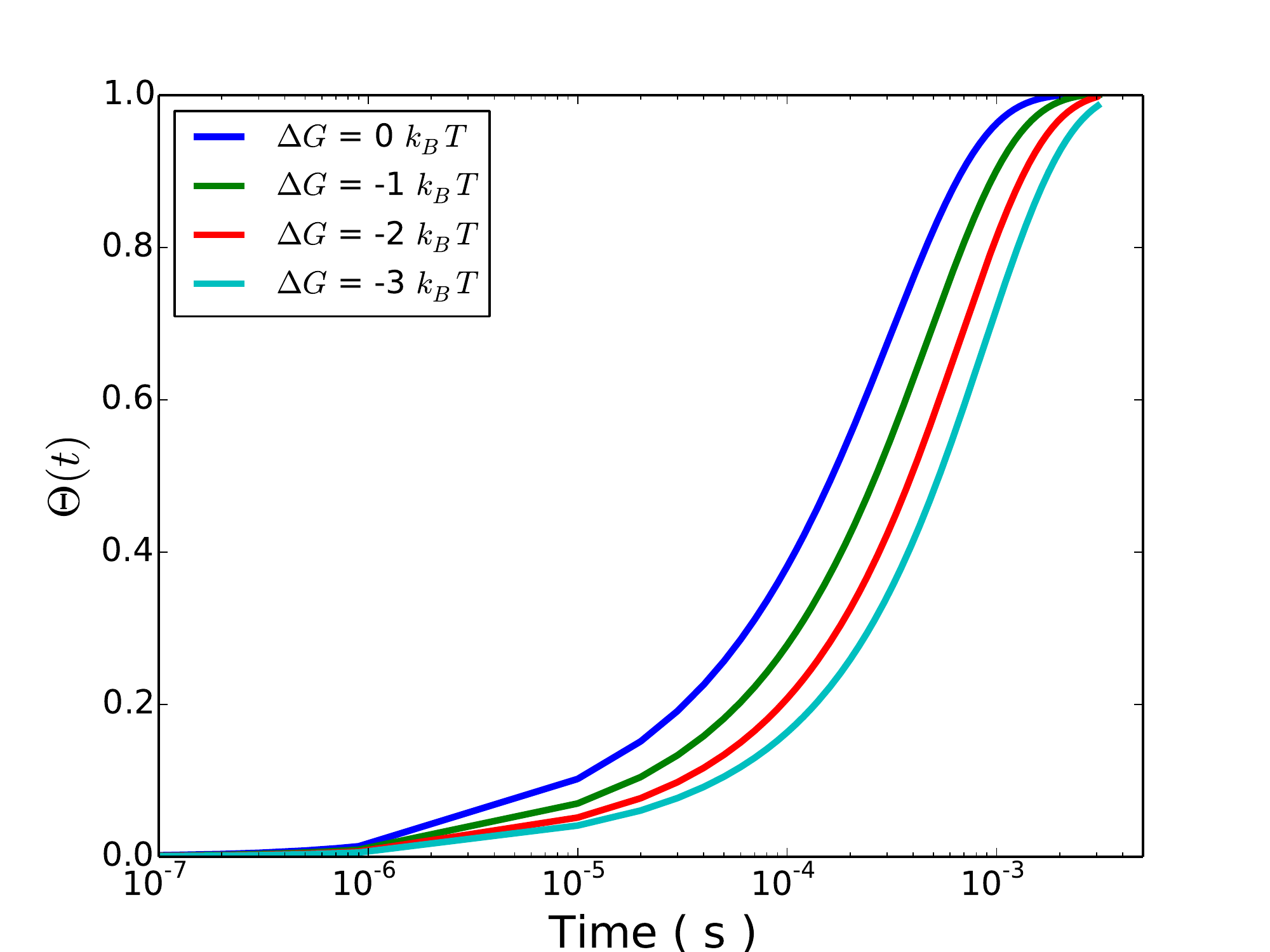}\label{fig:dG1}\\
\includegraphics[width=0.48\textwidth]{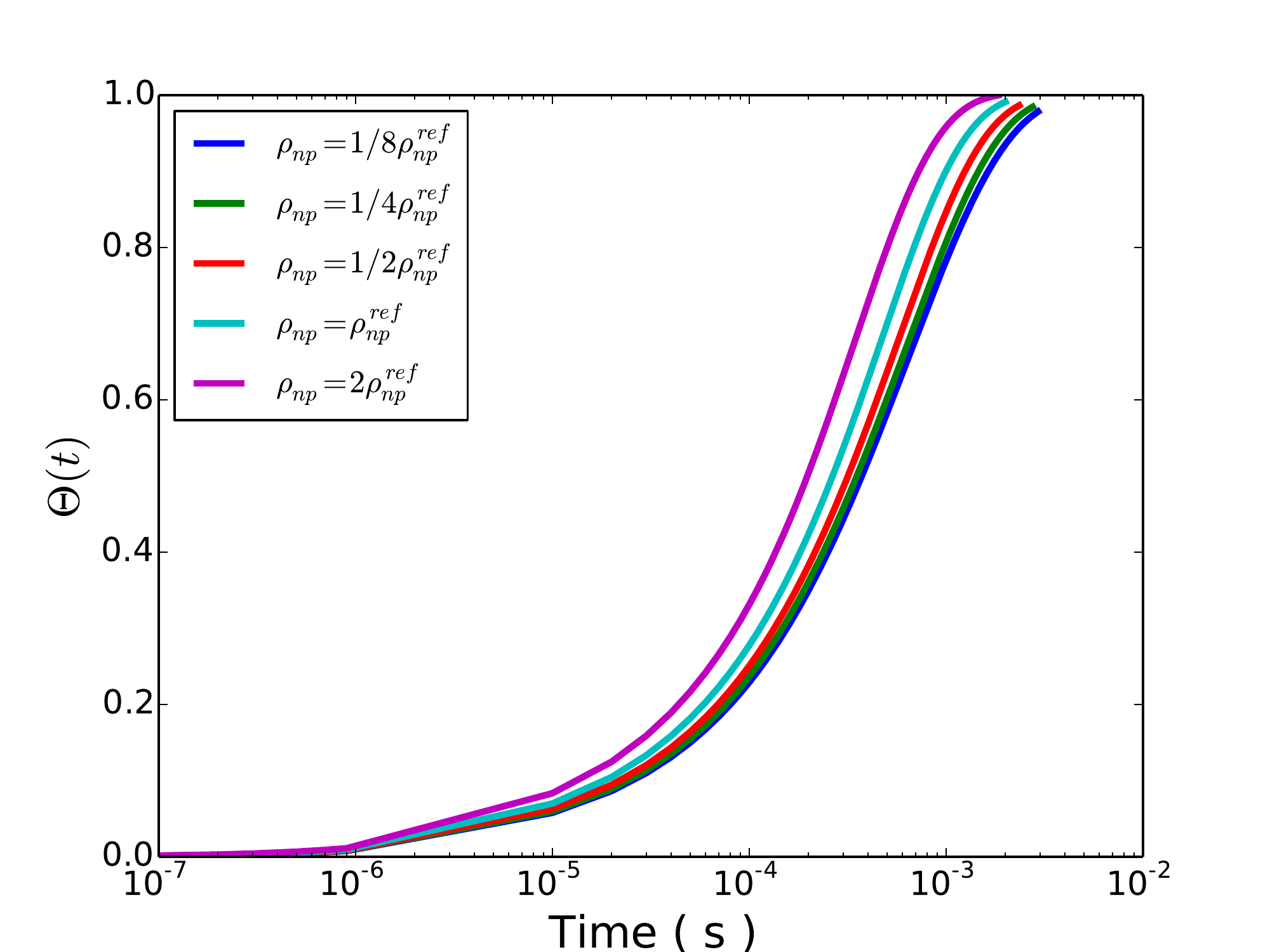}\label{fig:concentration1}
\includegraphics[width=0.48\textwidth]{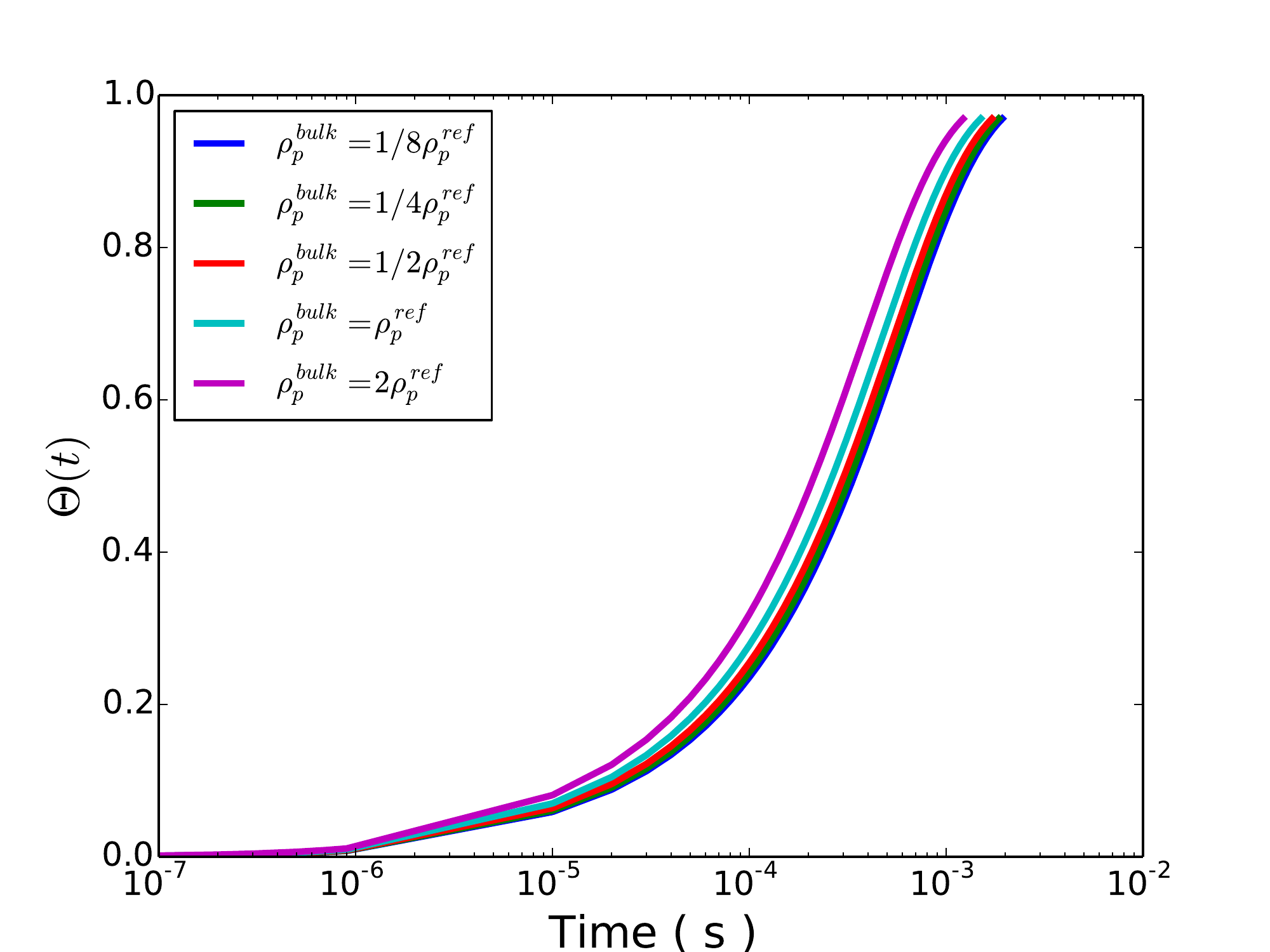}\label{fig:concentration_pro1}
\caption{Summary of simulations, loading $\Theta$ vs various parameters in our system. From top to bottom (left to write),
$\Theta$ is plotted a a function of protein's valence, intrinsic adsorption energy, nanoparticles' concentration and protein concentration.}   
\label{fig:compare-all}
\end{figure*}

\begin{figure*}[h]
\includegraphics[width=0.48\textwidth]{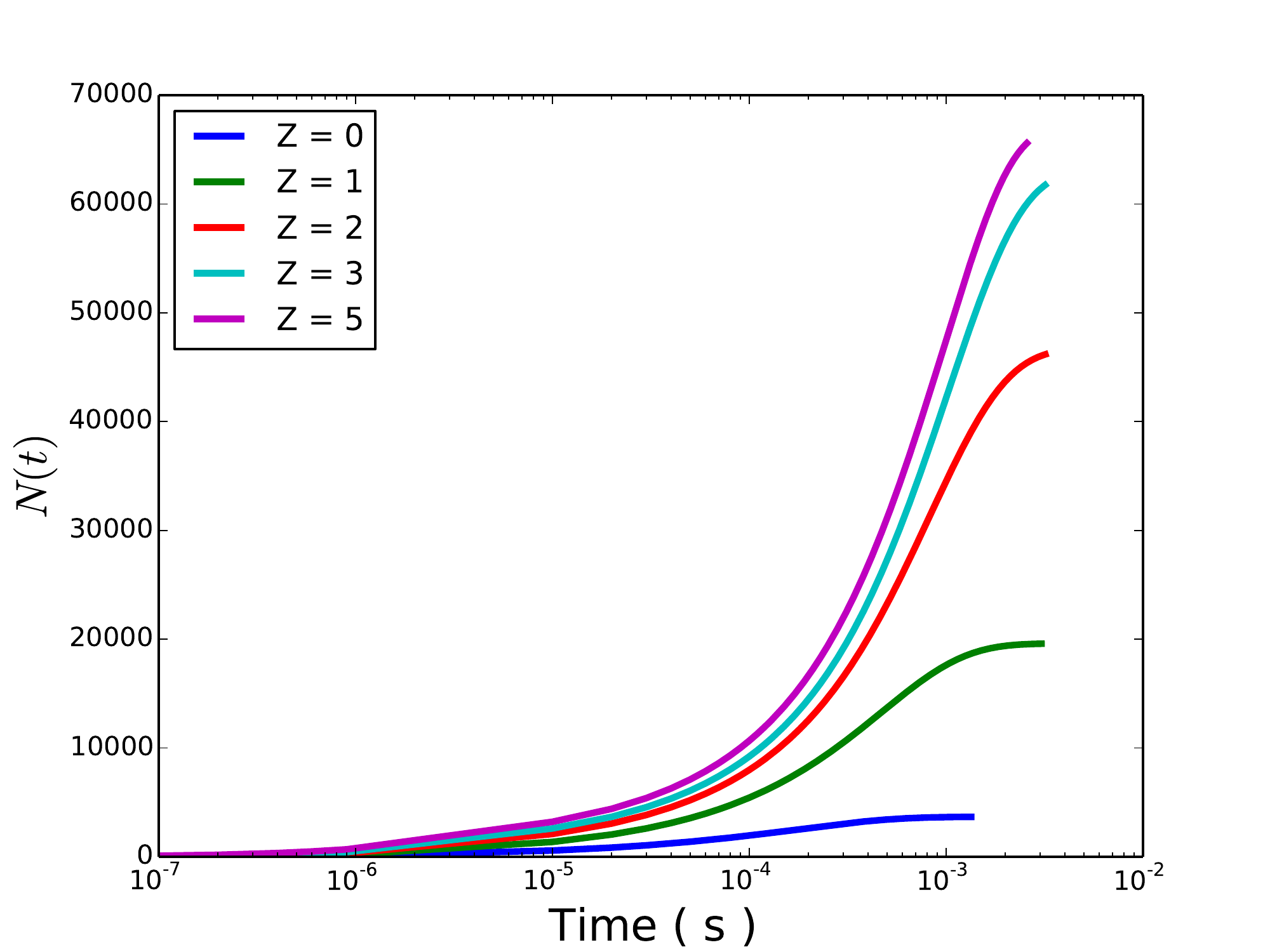}\label{fig:valence_unnorm1}
\includegraphics[width=0.48\textwidth]{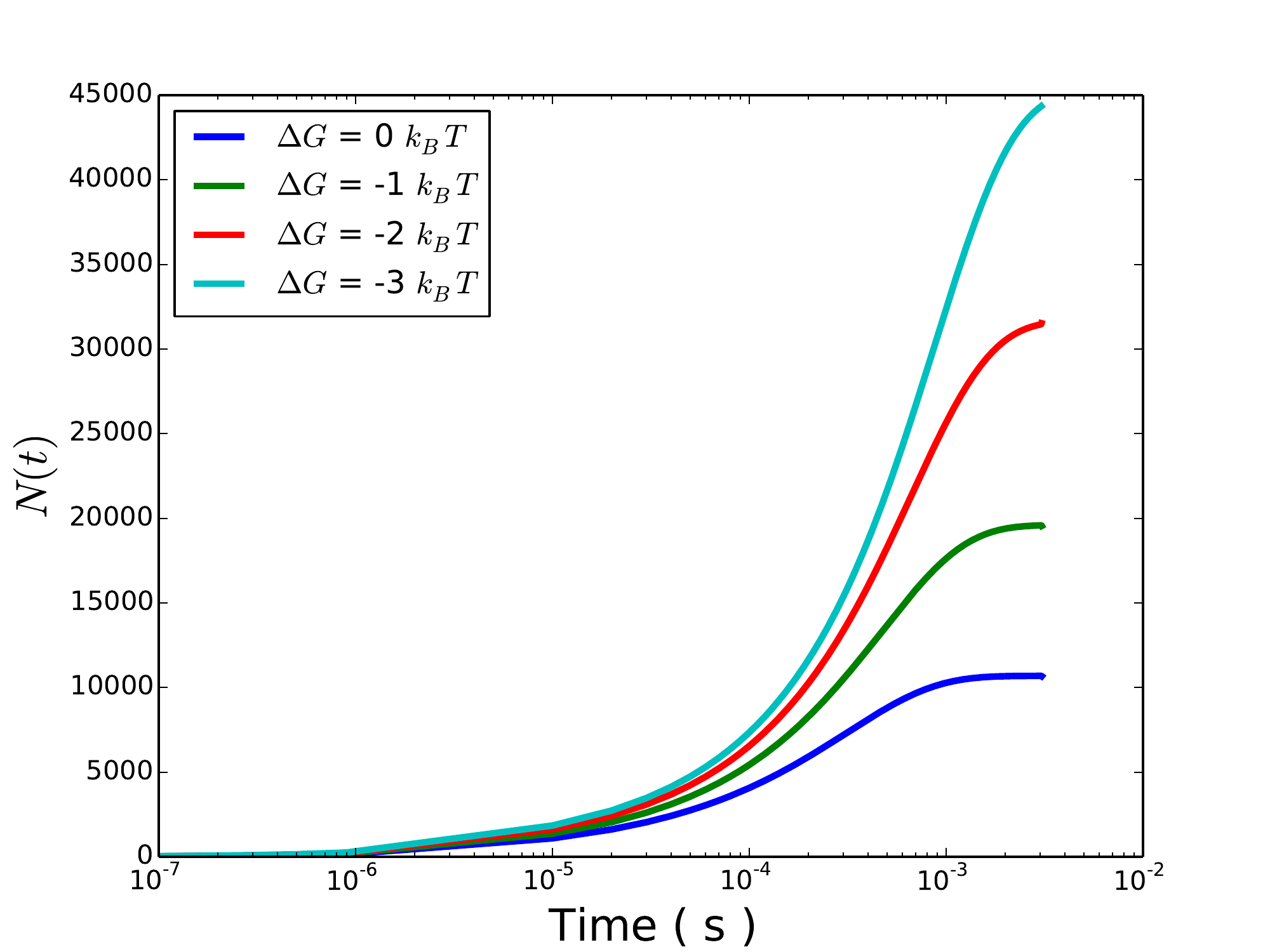}\label{fig:dG_unnorm1}\\
\includegraphics[width=0.48\textwidth]{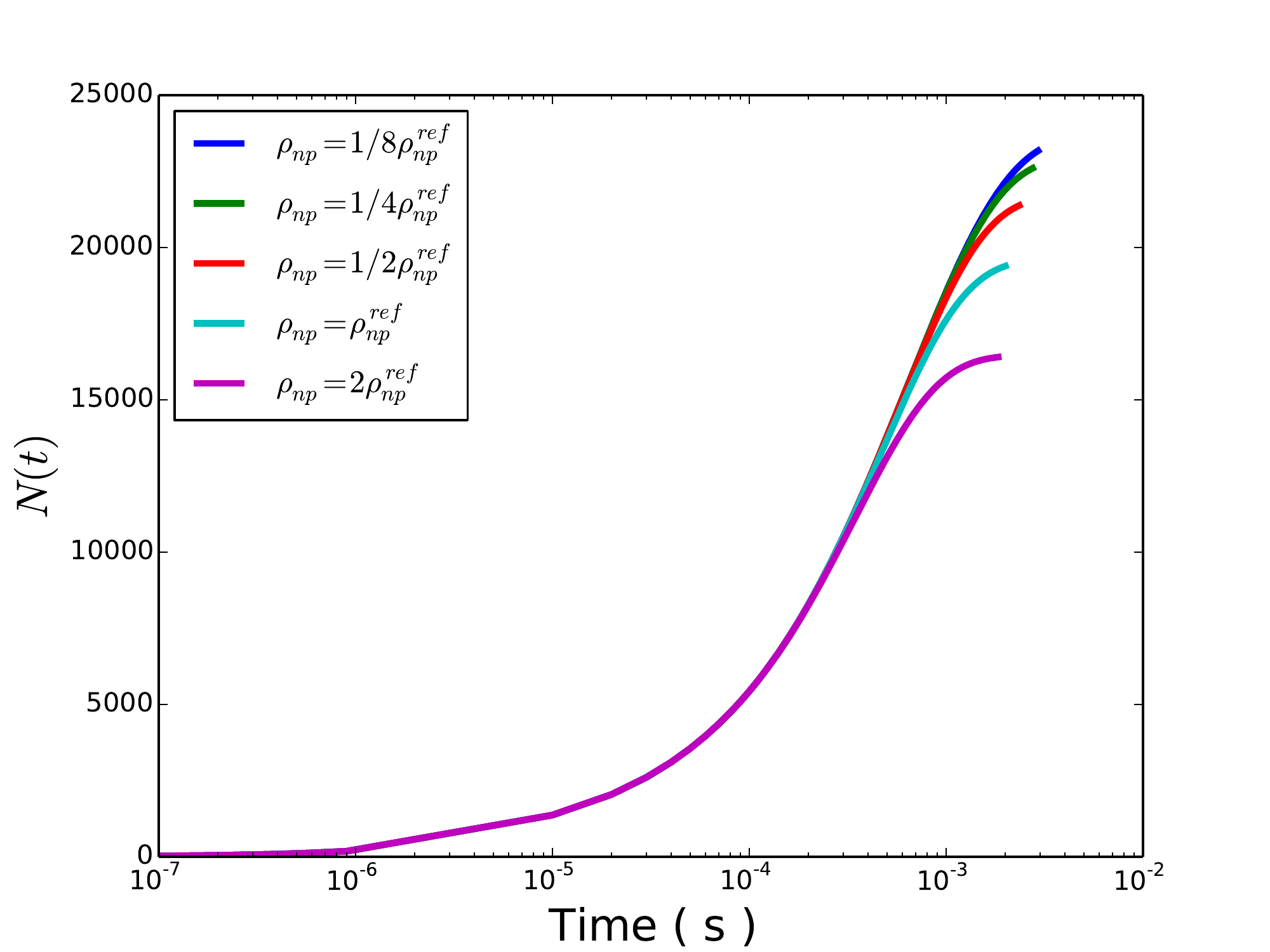}\label{fig:concentration_unnorm1}
\includegraphics[width=0.48\textwidth]{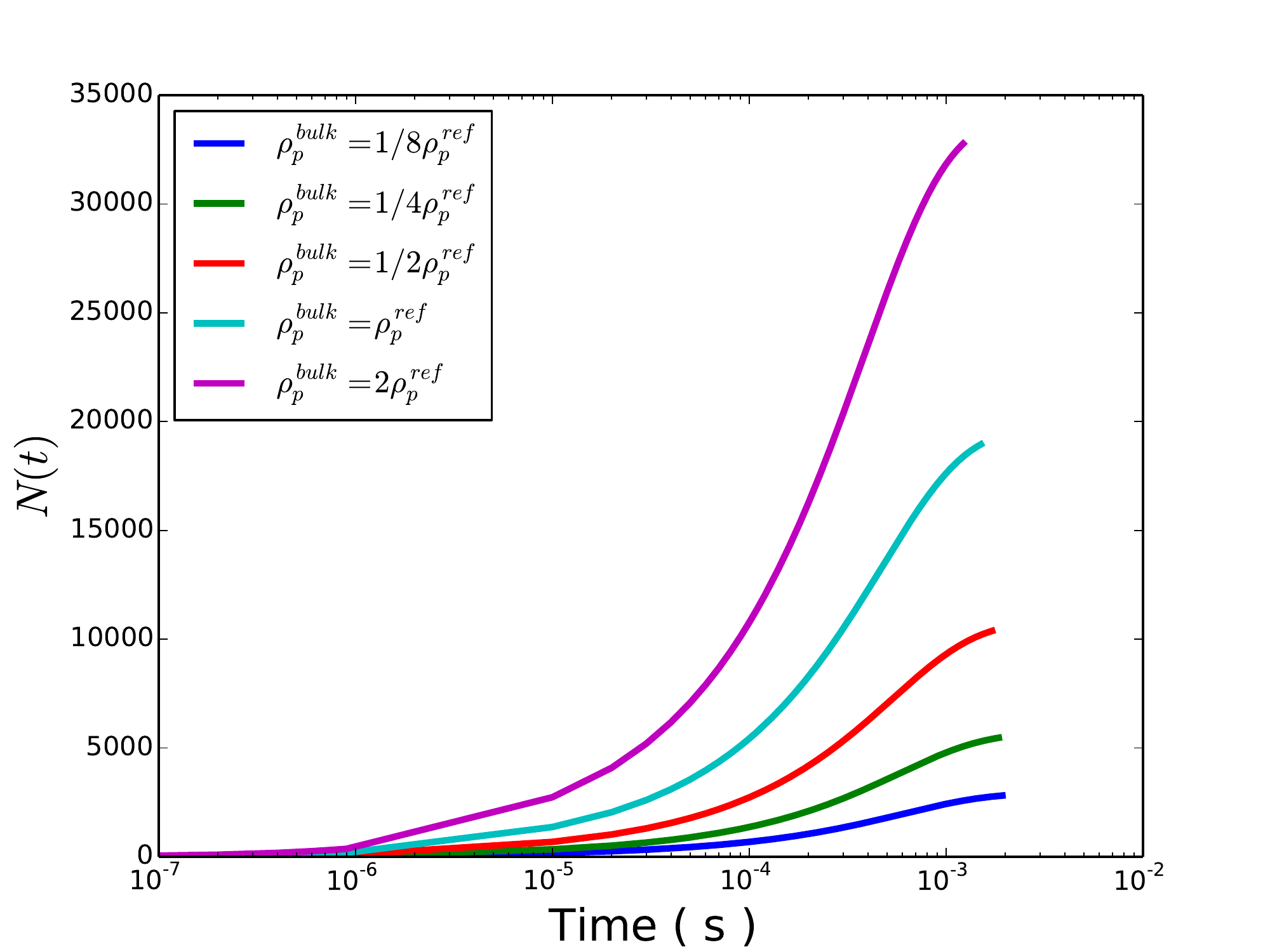}\label{fig:concentration_pro_unnorm1}
\caption{As in Fig.~\ref{fig:compare-all}, but here the un-normalised amount of adsorbed proteins is reported.}   
\label{fig:compare-all-unnorm}
\end{figure*}

\begin{table}
\begin{tabular}{l |r || c}
\hline
\hline
& $t_{1/2}~(\mu s)$ & Trend \\
\hline
\hline
Z & & \\
 0&  90 & \\
  1& 260  & \\
  2 &  510 & non-monotonous \\
   3 &  620 \\
     5   & 560   & \\
    \hline
$\rho_{p}^{bulk}/\rho_p^{ref}$& & \\
1/8 & 340 & \\
1/4  & 330 & \\
1/2   & 300 & decreasing \\
1    &  270 & \\
 2       & 210 & \\
\hline
$\rho_{np}/\rho_{np}^{ref}$& & \\
1/8 & 370  & \\
1/4  & 350 & \\
  1/2 & 320   & decreasing \\
    1 & 270 & \\
      2  & 200 & \\
\hline
$\mid \beta\Delta G^{ads} \mid$& & \\
0 &  160 & \\
 1 & 270 & \\
  2 & 390 & increasing \\
  3 & 530 & \\
\hline
\end{tabular}
\caption{Time to reach half the equilibrium loading $t_{1/2}$ for the various parameters combinations investigated
in our system. Note in particular that $t_{1/2}$ as a function of valence shows a peculiar non-monotonous behaviour,
possibly due to a maximum in the total amount of proteins adsorbed at equilibrium as a function of valence.
$\rho_p^{ref}$ and $\rho_{np}^{ref}$ are equal to $2.02\cdot 10^{-4}$~M and  $3.37\cdot 10^{-9}$~M, respectively.}
\label{tab:time}
\end{table}

The observed trends in Fig.~\ref{fig:compare-all} and Fig.~\ref{fig:compare-all-unnorm} can be 
rationalised in terms of two balancing mechanisms.
On the one side, higher driving forces, for example a lower $\Delta G^{ads}$ or higher 
protein concentration should lead to a faster kinetics, given that higher fluxes are expected. 
The same should happen for lower nanoparticle concentrations, for which the counter flux,
introduced via the boundary conditions that account for proteins being adsorbed by neighbouring particles, is reduced.
This is indeed the case, because at any one time the amount of adsorbed proteins is an increasing function 
of these driving forces, as can be  observed from the unnormalised adsorption profiles of Fig.~\ref{fig:compare-all-unnorm}.
On the other side, however, higher driving forces (with the only exception of the protein's valence,
which deserve a separate discussion later) also lead to a higher number of adsorbed proteins at equilibrium.\\
Clearly, if both the equilibrium number of proteins adsorbed and the average fluxes were linearly increasing 
functions of these driving forces, $\Theta(t )$ for different parameter values, i.e. Fig.~\ref{fig:compare-all} 
should collapse onto a single curve. Instead, a very different behaviour is observed.
In fact, it turns out that the increase in the total flux when higher driving forces are present 
is not always enough to compensate for the higher value of proteins that must be adsorbed to reach equilibrium, 
hence the loading dynamics can be slower. For example, loading as a function of increasing (in modulus) adsorption
energy becomes slower, whereas it is faster if we simply increase the initial bulk concentration of proteins, $\rho_{p}^{bulk}$, despite
in both ways we are increasing both the adsorption fluxes and $N(\infty)$.
However, although $N(\infty)$ as a function of $\rho_{p}^{bulk}$ grows more rapidly than for $\Delta G^{ads}$, its associated
flux increases even faster and the overall loading dynamics is actually faster and not slower for this latter case. 
As this example shows, the fact that both fluxes and equilibrium adsorption are 
highly non-linear functions of the control parameters implies that predictions based on simple arguments can be highly misleading, 
and one really has to solve the full equation of motion to rationalise these behaviours.
To make an even simpler example, let us just point out that for ideal diffusion the loading dynamics is not even a function of 
the bulk protein concentration, $\rho_{p}^{bulk}$.

An even stronger manifestation of non-linear behaviour can be observed in our system for the case of $\Theta(t)$ 
as a function of protein's charge $Z$. In this case, $t_{1/2}$ has a maximum for $Z=3$ and then decreases, a type of non-monotonic 
behaviour which would be difficult to predict without a full DDFT modelling. 
This maximum again arises since the total amount of adsorbed proteins at equilibrium $N(\infty)$ as a function of their charge rapidly 
saturates (see Fig.~\ref{fig:compare-all-unnorm} and compare the $Z=3$ and $Z=5$), whereas the thermodynamic force for adsorption does not
(at least until charge inversion of the loaded gel occurs).
Saturation is expected because of two competing effects.
On the one hand, when a protein of unlike charge absorbs the system decreases 
its energy by an amount $\mid Z V^{\mathrm{Don}} \mid$. 
However, $V^{\mathrm{Don}}$ is itself a function of the adsorbed charge, and becomes lower the higher the number of 
proteins in the gel. Hence, a maximum amount of adsorbed particles exists, when the adsorption of one more protein would effectively 
increase the total electrostatic potential felt in such a way that no-more energy is gained. Given the form of $V^{\mathrm{Don}}$ 
( Eq.~\ref{eq:donnan2} ), this is expected to happen earlier for proteins of higher charge.\\
We would like to stress the fact that it would be difficult to rationalise these effect looking purely at the loading dynamics $\Theta(t)$ and not at the
``raw'' quantity $N(t)$, since the latter typically shows a different behaviour. In particular, terms as ``fast'' or ``slow'' dynamics should be used
based on one or the other quantity in order to avoid confusion, especially when comparing different systems, like for example nanoparticles
of different size. In this regard, we notice that many analysis of experimental results are often based on $\Theta(t)$ alone, although in 
principle such techniques have access to the raw quantity as well.\\

What additional insights do these simulations offer regarding protein adsorption adsorption on nanogels? One thing to notice is that the parameters' range
scanned in this systematic study covers typical values observed for protein-nanogels system, and the timescales observed should
thus be indicative of those expected in realistic scenarios. In this regard, we would like to highlight the fact that here protein adsorption occurs on
timescales of a few milliseconds. Whereas this will depend on the exact concentration of both nanoparticles and proteins, it is nonetheless many
orders of magnitude faster than that observed in typical anti-fouling applications such as PEG-coated surfaces \cite{protein-critique}, or for bare nanoparticles
\cite{linse,riviere}.
Hence, it is reasonable to assume that in this system the protein's corona always reaches equilibrium with the local environment. This fact can have important
repercussions on large-scale models for farmacokinetics, since it would justify modelling the nanoparticles behaviour in the human body assuming the protein
corona (i.e. the nanoparticles "biological identity" \cite{nature-protein-corona2, lundqvist}) rapidly adapts to the changes in pH, protein and salt concentration found in different tissues
(given that transport between different parts of the body of these nanoparticles by either diffusion or convection through the blood-stream occurs on timescales a few 
orders of magnitude higher).
This is clearly not the same behaviour one can assume to describe, for example, protein induced degradation in a biomedical implant, since the protein adsorption
kinetics in this case will necessarily play a much more important role given the long times required to achieve equilibrium.\\
These conclusions might be challenged when considering the case of competitive protein adsorption when multiple types are present, which will be studied 
in a future publication.

\section{Conclusions \label{sec:conclusion}}
In this paper, we presented a theoretical model based on DDFT to describe protein 
adsorption on charged, polymer-coated nanoparticles. Compared to simpler descriptions 
of the kinetics such as models based on ideal diffusion or Langmuir-type kinetics, 
DDFT offers a natural and very general framework to include in a controlled manner the 
effect of all possible interactions within the system, and to separately study their effect.\\*
Here, we concentrated on including those effects which proved to be useful to rationalise
the adsorption thermodynamics in the system, and separate interactions into non-specific, 
global electrostatic interactions as captured by the concept of the Donnan potential and Born energy,
and protein-specific, intrinsic effect such as those arising from hydrophobic interactions and 
excluded volume effects \cite{joe-langmuir}.\\
The model is constructed so that once the intrinsic adsorption energy is obtained by fitting calorimetric curves 
probing the thermodynamics of protein adsorption in the system, the kinetics can be described with no additional parameter.
Using such a procedure, we are able to reproduce on a semi-quantitative level the observed experimental 
loading kinetics of Lysozyme on PNIPAM coated nanogels.\\
Finally, we presented a parametric exploration of the model, where we studied 
the variation in the loading kinetics for various quantities of interest, such as protein's valence and intrinsic adsorption
energy, as well as their concentration and that of the nanogels in solution.
Curiously, in all cases the timescale for protein adsorption is on the millisecond scale, suggesting
fast equilibration of the protein corona with the local environment for typical settings where nanoparticles are used,
for example, for drug delivery.\\
Before we conclude, we have a last remark. 
Although we applied it here for the case of a single-component system to present its main feature, the 
model can be easily extended to the case of multi-component systems, where possible cooperative and/or 
competitive adsorption effects are expected, giving rise to a peculiar, non-monotonic dynamics in the 
adsorption profiles such as those observed in the so-called "Vroman effect" \cite{vroman,vroman-effect-review}. 
Modelling of such phenomena are currently under investigation, and will be the presented in future publications.

\section{Acknowledgements}
S.A-U and J.D acknowledge funding from the Alexander von Humboldt (AvH) Foundation via a Post-Doctoral Research
Fellowship. All authors acknowledge support from the Helmholtz Virtual Institute (HVI) "Multifunctional Materials in Medicine" 
(Berlin and Teltow), Germany.

\providecommand*{\mcitethebibliography}{\thebibliography}
\csname @ifundefined\endcsname{endmcitethebibliography}
{\let\endmcitethebibliography\endthebibliography}{}

\end{document}